\documentclass[11pt]{article}
\usepackage[margin=1in]{geometry}
\usepackage{amsmath,amssymb,bm}
\usepackage{booktabs}
\usepackage{graphicx}
\usepackage{array}
\usepackage{colortbl}
\usepackage{xcolor}
\usepackage{microtype}
\usepackage{enumitem}
\usepackage{xcolor}
\usepackage{float}
\usepackage[authoryear,round]{natbib}
\newcommand{\vstrut}{\rule[-2ex]{0pt}{5ex}}
\usepackage[width=0.8\textwidth]{caption}
\newcommand{\E}{\mathbb{E}}
\newcommand{\Var}{\operatorname{Var}}

\begin{document}
\sloppy

\begin{center}

{\Large \textbf{How deterministic trajectories and their fluctuations} }

\medskip

{\Large \textbf{underlie allele-frequency statistics in the}}

\medskip

{\Large \textbf{strong-selection regime}}

\medskip

\bigskip

David Waxman

\medskip

Centre for Computational Systems Biology

Institute of Science and Technology for Brain Inspired Intelligence

Fudan University, 220 Handan Road, Shanghai 200433, PRC

\bigskip

\bigskip

{\large{\textbf{Abstract}}}
\end{center}

 We determine statistics of allele-frequency trajectories across 
 replicate populations. Working under the diffusion approximation, we define $R = 2N_e|s|$, 
 where $N_e$ is the effective population size and $s$ is the additive selection 
 coefficient of a focal allele. We focus on strong selection, corresponding to 
 $R \gg 1$. Strong selection, however, does not require large selection coefficients: 
 in populations of realistic size, it can readily arise from small selection coefficients, 
 which commonly occur. Our analysis concerns a biallelic locus over timescales on 
 which mutation can be neglected. Every allele-frequency trajectory then belongs to 
 one of two absorbing classes: fixation or loss.

When $R$ is large, individual allele-frequency trajectories can appear nearly deterministic. 
This does not imply that an approximation of allele-frequency statistics 
can be based on a single deterministic trajectory. For a beneficial allele, the deterministic 
trajectory approaches fixation, yet a non-negligible fraction of stochastic trajectories 
in replicate populations may be lost, particularly when the allele starts at low frequency. 
Ignoring trajectories that achieve loss can produce severely erroneous statistics, for example, 
the mean allele frequency can differ drastically from the deterministic trajectory.  
We address this problem by conditioning separately on the eventual loss and the eventual fixation 
of the focal allele. This yields an outcome-conditioned representation of an allele-frequency statistic.
In the strong-selection regime, we apply a small-noise (small-drift) expansion that reasonably 
captures transient behaviours of statistics while recovering their exact long-time limits. 
We illustrate results with numerical examples for the mean allele frequency and its variance. 

The biological interpretation of this work is that, even under strong selection, both trajectory 
types, those that eventually reach fixation and those that are eventually lost, are of importance. 
Averaging statistics over both outcomes forces both trajectory types into the analysis, 
thereby retaining trajectories that a single deterministic approximation would omit. Furthermore, 
under strong selection, conditioning on fixation or loss leads trajectories of both type to be primarily 
deterministic, with small corrections due to fluctuations. This provides a simple framework for calculating 
time-dependent allele-frequency statistics relevant to experimental evolution, longitudinal 
population-genomic data, and the dynamics of alleles driving selective sweeps.

\newpage
\section*{Introduction}

Allele-frequency dynamics are a central component of population genetics
because they connect evolutionary forces acting at individual loci with the
distribution of genetic variation observed across populations and through
time \citep{EwensBook}.  The Wright-Fisher model and its diffusion
approximation provide a standard framework for analysing selection and genetic
drift \citep{Fisher1922,Wright1931,Wright1945,Kimura1955}.  Much of the
classical theory and many practical applications focus on neutral or weakly
selected alleles.  At the same time, selection can be strong relative to drift
in biologically relevant populations, and there is increasing interest in
inference across a broad range of selection strengths
\citep{Gillespie1983,Lacerda2014PopulationGI,Tataru}.

Let $s$ denote the selection coefficient of a focal allele at a biallelic locus, 
with selection assumed additive (no dominance). Throughout this work, we assume, as is common, 
that $s$ is small ($|s|\ll1$). Whenever we say selection is `weak' or  `strong', 
the comparison is always with the effects of genetic drift.

With $N_e$ the variance effective population size, we introduce the composite parameter
\begin{equation}
 R=2N_e|s|
 \label{eq:R}
\end{equation}
which we will call the `scaled strength of selection'. 
Under the diffusion approximation, $R$ quantifies the strength of selection 
relative to random genetic drift. Thus $R\ll1$ corresponds to selection that is weak
relative to drift, whereas $R\gg1$ defines the strong-selection regime studied
here.  Importantly, $R \gg 1$ does not require that $s$ itself be large, and 
is fully compatible with $s$ being small.  For
example, $N_e=10^4$ and $s=\pm 10^{-3}$ gives $R=20$, while
$N_e=10^5$ and the same selection coefficients give $R=200$. 

Under strong selection, the characteristic timescale over which the frequency
of the focal allele changes is $\sim 1/|s|$, at least away from the
boundaries. Thus when $s$ is small, as we assume, ``selection being strong''
does not preclude the frequency from changing over hundreds or thousands of
generations.

\subsubsection*{Strong selection does not imply dominance of a single
deterministic trajectory}

A natural intuition, when $R$ is large,  is that stochastic trajectories
should lie close to the deterministic trajectory that is generated purely by selection.
That intuition is useful but incomplete when the quantity of interest is a
statistic calculated across replicate populations.

Consider an allele with $s>0$, i.e., a beneficial allele. 
The deterministic trajectory 
that starts at frequency $y$ at time $t=0$ is\footnote{The deterministic frequency at time $t$, written $X_{t}^{(\mathrm{det)}}$, satisfies the drift-free equation ${dX_{t}^{(\mathrm{det)}}/dt=sX_{t}^{(\mathrm{det)}}(1-X_{t}^{(\mathrm{det)}})}$.} 
\begin{equation}
X_{t}^{(\mathrm{det)}}=
 \frac{y}{y+(1-y)e^{-st}}.    \label{X deterministic}
\end{equation}
This satisfies $X_{t}^{(\mathrm{det)}}\to1$ as $t\to\infty$.  If one uses this trajectory
as the leading approximation to the mean allele-frequency then the
long-time value is predicted to approach $1$.
The exact stochastic dynamics of the Wright-Fisher model leads to a qualitatively 
different behaviour. Even a beneficial allele can be lost, and when its initial frequency 
is low, the probability of loss can be substantial. When this is the case, the long-time 
value of the mean allele frequency will be substantially smaller than $1$.  

The issue is not simply that the stochastic trajectory fluctuates around a deterministic path. 
The key point is that trajectories fall into two qualitatively distinct classes, those that 
eventually fix and those that are eventually lost, and that population-level allele-frequency 
statistics depend on both.

Thus even when selection is strong relative to drift, approximating such statistics,
using just the deterministic trajectory, can fail badly, and both absorbing
outcomes, fixation and loss, generally have to be incorporated into the
analysis. This leads to the central question of this paper.
\begin{quote}
\emph{When selection is strong relative to genetic drift, how do fixation and loss
combine to shape population-level statistics of the allele frequency?}
\end{quote}
Our analysis, developed through the lens of conditioning, leads to two main findings. 
First, strong selection suppresses stochastic fluctuations within each of the two classes 
of trajectories, defined by whether they eventually reach fixation or loss. Furthermore, 
each class is associated with a distinct deterministic trajectory.

Second, at long times, the contributions of these two deterministic trajectories 
recover population-level statistics. At shorter times, however, stochastic fluctuations 
may transiently dominate the variance, while making a comparatively small contribution to the mean.

\begin{figure}[H]
    \centering
    \includegraphics[width=1\linewidth]{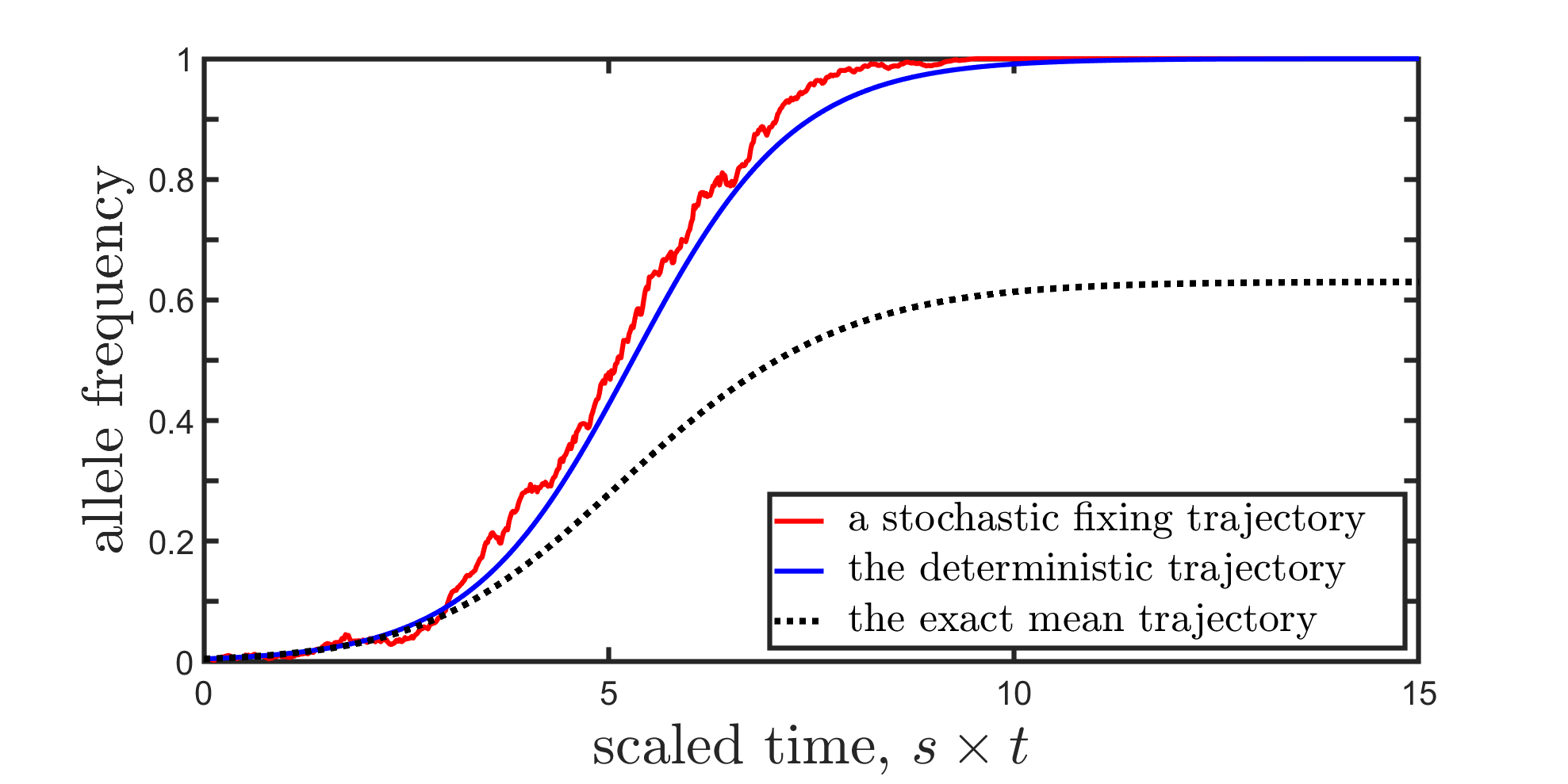}
    \caption{\textbf{Three distinct allele-frequency trajectories.} 
    The figure applies to a beneficial allele ($s>0$) in a randomly mating population of diploid hermaphrodites, where effective population size equals census population size.
    We plot three allele-frequency trajectories over time: (i) a stochastic trajectory that 
    arises from the exact stochastic dynamics and which achieves fixation; (ii) the deterministic 
    trajectory of Eq. (\ref{X deterministic}), which approaches $1$ as time proceeds; and 
    (iii) the exact mean allele frequency that arises from the Wright-Fisher model. 
    For the figure, the initial frequency is $y=5 \times 10^{-3}$, the effective population size is
    $N_e=3 \times 10^3$ and the selection coefficient is $s\simeq 1.67 \times 10^{-2}$, corresponding
    to a scaled strength of selection of $R=100$. The fixation probability is approximately 0.63.}
    \label{fig:1}
\end{figure}

Figure \ref{fig:1} illustrates that stochastic and  deterministic trajectories can be somewhat similar.
However, in the case illustrated, these trajectories differ substantially, at long times, from the exact 
mean allele frequency, which represents an average over many replicate populations. The exact mean allele frequency 
is derived (along with other statistics),  not only from the set of trajectories that achieve fixation, 
but also from the set that achieve loss. In Figure \ref{fig:1}, most trajectories
which eventually achieve loss would not, if plotted, be discernible, because they start at a low 
initial frequency, and most never rise to an appreciable frequency. 
However, trajectories which eventually achieve loss can strongly influence statistics that 
are calculated across replicate populations. For the parameter values used in 
Figure \ref{fig:1}  the trajectories which eventually achieve loss make up close to
$37\%$ of replicate-population trajectories, and lead to a mean allele 
frequency, at long times, that is close to $0.63$.

\subsection*{Biological significance}

The distinction between the contribution of a single deterministic trajectory 
to population-level statistics, and the contributions of trajectories in the
two absorbing classes, is relevant whenever allele-frequency data are interpreted
across replicate populations or across repeated evolutionary histories.
Examples include experimental evolution, longitudinal sampling of natural
populations, and rapidly adapting microbial or viral populations.  In such
settings, the observation or knowledge that an allele is strongly favoured does not by
itself imply that every replicate population follows the same trajectory or
reaches the same endpoint.  Early stochastic loss and subsequent deterministic
increase are distinct components of the process.

The distinction is also relevant to population-genomic studies of selective
sweeps.  Classic sweep theory predicts characteristic changes in linked
neutral variation, including reductions in local diversity, distortions of
the site-frequency spectrum, and extended haplotype structure
\citep{MaynardSmithHaigh,Nielsen,Sabeti}.  Genome-wide scans
therefore often use allele-frequency and haplotype patterns as evidence for
recent positive selection.  The present analysis does not replace such sweep
models, because it treats a single biallelic locus and neglects linkage.
Instead, it provides a complementary way of thinking about the trajectory of
the selected allele itself: the observed time-dependent distribution can
contain contributions from alleles that are destined to increase to fixation and
alleles that are ultimately eliminated.

This distinction is particularly relevant for partial sweeps, temporally
sampled populations, and analyses in which the selected allele is observed
before its ultimate fate is known.  The framework developed here therefore
provides a possible building block for future population-genomic models rather
than a complete sweep-detection method.

\subsection*{Relation to existing approaches}

The classical diffusion framework describes the full stochastic process
\citep{Kimura1955}.  Fixation probabilities and their strong-selection
limits are also well established \citep{Haldane1927,Kimura62}.  Conditional
diffusion methods have long provided a way of describing paths given an
eventual absorbing outcome \citep{Ewens1973}, and conditioning on fixation
has been used to obtain the characteristic trajectory of a fixing allele
\citep{ZhaoLascouxOverall}.  Strong-selection analyses have also considered finite
populations and strong selection from related perspectives
 \citep{Gillespie1983,Balick2023}.

The purpose of this work is different. We construct time-dependent statistics
by retaining \textit{both} absorbing outcomes and then develop a small-noise
expansion within each outcome-conditioned class. The resulting approximation
differs from a single-trajectory treatment in three respects: it retains the
relative weights of loss and fixation, it involves two deterministic
trajectories rather than one, and it includes stochastic fluctuations around
each deterministic trajectory.

The contribution here is not conditioning itself, but a systematic approach
to time-dependent statistics in the strong-selection regime. Within each
outcome-conditioned set, the approach uses a deterministic trajectory with
fluctuation corrections.

\section*{Biological model}

We consider a Wright-Fisher model \citep{Fisher1922,Wright1931} for a finite population of 
diploid hermaphroditic organisms, where generations are discrete, and in each
generation, $N$ adults are maintained in the population. 

A generation starts with adults who randomly pair and produce offspring.  
We assume there is no migration, and over the timescales considered 
we neglect mutation. 

After reproduction, the adults die, leaving the offspring, who are subject 
to viability selection at a single biallelic locus. We write the focal allele 
as $A$, and the alternative allele as $B$. The relative fitnesses of the three genotypes 
are taken to be
\begin{equation*}
     AA:\enspace 1+2s,\qquad AB:\enspace 1+s,\qquad BB:\enspace 1
\end{equation*}
i.e., selection is additive.
%We can think of $s$ as a selection coefficient associated with the $A$ allele. 
A non-selective thinning process then reduces the number of individuals in the population to $N$, 
who then constitute the adults of the next generation. 

The assumption of neglecting mutation over the timescales considered is
key to this work: it results in the allele frequency having two absorbing outcomes, namely
fixation and loss.  The framework is therefore aimed at the establishment and fate of
a focal allele for a finite evolutionary episode, rather than over longer timescales
where mutation, selection, and drift co-occur.

\section*{Diffusion approximation}

Adopting the diffusion approximation for the analysis, the random frequency of the focal allele at time $t$, written $X_t$, 
changes over the time interval $t$ to $t+dt$ according to
\begin{equation}
 dX_t=sX_t(1-X_t)\,dt+
 \sqrt{\frac{X_t(1-X_t)}{2N_e}}\,dW_t
 \quad \text{with } X_0=y.
 \label{X SDE}
\end{equation}
The first term on the right-hand side represents the deterministic contribution 
to the change in $X_{t}$ due to natural selection, while the second term 
describes random genetic drift. In this term, $N_e$ denotes the \textit{effective population size},
which typically replaces the census population size ($N$) due to deviations from an 
ideal Wright-Fisher model \citep{Wright1931},
while $dW_{t}$ denotes an uncorrelated random fluctuation \citep{Tuckwell}. Equation (\ref{X SDE}) is equivalent 
to the diffusion equation that governs the distribution of $X_t$  \citep{Kimura1955}.

To proceed, we carry out a transformation of the time in Eq. (\ref{X SDE}). We
define
\begin{equation}
\tau=|s|t,\qquad Y_{\tau}=X_{\tau/|s|},\qquad\sigma=\operatorname{sign}%
(s),\qquad\varepsilon=1/R.   \label{tau Y sigma eps}
\end{equation}
The quantity $\tau$ is the transformed time. The quantity $Y_{\tau}$ is the time-transformed
random frequency and obeys
\begin{equation}
 dY_\tau=\sigma Y_\tau(1-Y_\tau)\,d\tau+\sqrt{\varepsilon}\cdot\sqrt{Y_\tau(1-Y_\tau)}\,dW_\tau
 \quad\text{with } Y_0=y
 \label{Y SDE}
\end{equation}
 - see Appendix \ref{transformation SDE appendix} for details.

Strong selection corresponds to $\varepsilon \ll 1$. Given the presence of $\sqrt{\varepsilon}$ 
in Eq. (\ref{Y SDE}) a natural step is to  expand $Y_\tau$ in powers of $\sqrt{\varepsilon}$ and 
write $Y_{\tau}=Y_{\tau,0}+\sqrt{\varepsilon} \,Y_{\tau,1}+\varepsilon \, Y_{\tau
,2}+\ldots$\quad. However this procedure leads to a single deterministic 
trajectory along with small fluctuations of order  $\sqrt{\varepsilon}$  
around this trajectory. For $s>0$ the deterministic trajectory is equivalent to that in Figure \ref{fig:1} (blue curve).
Generally, this procedure proves inadequate for the reasons already stated, that 
trajectories which eventually achieve loss can strongly influence statistics, 
and we proceed differently.

\section*{Statistics incorporating both absorbing outcomes}

Let $q(Y_{\tau})$ represent a function of the random frequency $Y_{\tau}$. The
basic statistic we consider is an expectation of $q(Y_{\tau})$ that  we write as 
\begin{equation}
Q(\tau)=\E_{y}[q(Y_{\tau})].   \label{Q def}
\end{equation}
%Here, $\E_{y}[q(Y_{\tau})]$ is the expectation of $q(Y_{\tau})$, 
The notation $\E_y[q(Y_\tau)]$ denotes an expectation taken with respect to
the distribution of the random frequency $Y_\tau$, for an initial frequency of $y$ (i.e., $Y_0 = y$).
We adopt the convention that the expectation of a function of \textit{any} 
random frequency, say $Z_\tau$, is taken with respect to the distribution of $Z_\tau$.

To calculate the mean allele frequency, we use $q(x)=x$. For statistics
such as the variance or the mean heterozygosity, we use two forms of $q(x)$, namely $q(x)=x$ and $q(x)=x^2$,
and combine the resulting $Q$'s to obtain the required quantity.

\subsubsection*{Conditioning}
Let us define
\begin{equation}
\left.
\begin{array}
[c]{l}%
\pi^{(0)}(y)=\text{probability of eventual loss of the }A\text{ allele,
for initial frequency }y\\[1.5ex]
\pi^{(1)}(y)=\text{probability of eventual fixation of the }A\text{ allele,
for initial frequency }y.%
\end{array}
\right\}
\end{equation}
These satisfy $\pi^{(0)}(y)+\pi^{(1)}(y)=1$.

For the form of selection adopted, the diffusion approximation
gives
\begin{equation}
\pi^{(1)}(y)=\frac{1-e^{-2\sigma Ry}}{1-e^{-2\sigma R}}\label{eq:fixprob}%
\end{equation}
\citep{Kimura62}. 

Then, with $loss$ and $fix$ shorthand for the eventual occurrence of loss and
fixation, respectively, the law of total expectations gives
\begin{align}
Q(\tau)  & =\E_{y}[q(Y_{\tau})]\nonumber\\[1.5ex]
%& \nonumber\\
& =\E_{y}[q(Y_{\tau})|loss]\cdot\pi^{(0)}(y)+\E_{y}[q(Y_{\tau})|fix]\cdot
\pi^{(1)}(y).\label{eq:conditioning}%
\end{align}
Equation~\eqref{eq:conditioning} forces the statistic $Q(\tau)$
to derive contributions from both classes of trajectory, those which achieve loss 
and those which achieve fixation.

\subsubsection*{Modified random frequencies}

The expectation $\E_{y}[q(Y_{\tau})|loss]$ appearing in Eq. (\ref{eq:conditioning})
involves an average over all trajectories of Eq.~(\ref{Y SDE}) that start at frequency $y$ at $\tau=0$
and eventually achieve loss. Such an expectation does not admit an obvious direct approximation.
However, the set of trajectories that are conditioned on loss 
is equivalent to a set of \textit{unconditioned trajectories}, and quantities based on
unconditioned trajectories are amenable to approximation, as we will shortly show.
The unconditioned trajectories arise
from an equation similar 
to  Eq.~(\ref{Y SDE}), but with a \textit{modified form of selection} 
\citep{Ewens1973,ZhaoLascouxOverall}. Similarly, the set of all trajectories used 
in $\E_{y}[q(Y_{\tau})|fix]$ are also equivalent to a set of unconditioned trajectories, 
 but with a different modified form of selection.  
We write
\begin{equation}
\left.
\begin{array}
[c]{l}%
Z_{\tau}^{(0)}=\text{a modified version of }Y_{\tau}\text{ that eventually
achieves }0\text{ (loss)}\\[1.5ex]
Z_{\tau}^{(1)}=\text{a modified version of }Y_{\tau}\text{ that eventually
achieves }1\text{ (fixation).}%
\end{array}
\right\}
\end{equation}
The equations that $Z_{\tau}^{(0)}$ and $Z_{\tau}^{(1)}$ obey can be jointly written as
\begin{equation}
dZ_{\tau}^{(a)}=f^{(a)}\big(Z_{\tau}^{(a)}\big)\,d\tau+\sqrt{\varepsilon}\cdot
\sqrt{Z_{\tau}^{(a)}\big(1-Z_{\tau}^{(a)}\big)}\,dW_{\tau}\qquad\text{for }a=0,1\label{Z^a SDE}%
\end{equation}
where, compared with Eq. (\ref{Y SDE}), the forms of selection are modified
and given by
\begin{equation}
\left.
\begin{array}
[c]{rcl}%
f^{(0)}\left(  z\right)   & = & -\coth(R(1-z))\,z(1-z)\\[1ex]%
f^{(1)}(z) & = & \coth(Rz)\,z(1-z)%\\[1ex]%
%g(z) & = & \sqrt{z(1-z)}%
\end{array}
\right\}  \label{ff}%
\end{equation}
- see Appendix \ref{modified dynamics appendix} for details. 
Furthermore, the 
conditional expectations of the original frequency, $Y_{\tau}$, that are present in 
Eq. (\ref{eq:conditioning}),  are equal to
\textit{unconditional} expectations of the modified frequencies $Z_{\tau}^{(0)}$ and $Z_{\tau}^{(1)}$:
\begin{equation}
\mathbf{E}_{y}\big[q(Y_{\tau})\,|\,loss\big]=\mathbf{E}_{y}\big[q(Z_{\tau
}^{(0)})\big]\qquad\text{and}\qquad\mathbf{E}_{y}\big[q(Y_{\tau}%
)\,|\,fix\big]=\mathbf{E}_{y}\big[q(Z_{\tau}^{(1)})\big].\label{Eq loss fix}%
\end{equation}
These results, when used in Eq. (\ref{eq:conditioning}), 
allow us to write  the statistic 
$Q(\tau)=\mathbf{E}_{y}\left[  q(Y_{\tau})\right]$ in terms of unconditioned
expectations of the modified frequencies:
\begin{equation}
Q(\tau) =\mathbf{E}_{y}\big[q(Z_{\tau}^{(0)})\big]\cdot\pi^{(0)}(y)+\mathbf{E}
_{y}\big[q(Z_{\tau}^{(1)})\big]\cdot\pi^{(1)}(y).\label{Representation}
\end{equation}

\subsubsection*{Small-noise expansion}

The representation of $Q(\tau)$ in Eq. (\ref{Representation}) is suitable for
a \textit{small-noise expansion} \citep{Gardiner2009}; the modified
frequencies $Z_{\tau}^{(0)}$ and $Z_{\tau}^{(1)}$ are unencumbered by
conditioning, obey Eq. (\ref{Z^a SDE}), and each has only one accessible
absorbing value. Using $\sqrt{\varepsilon}$ as a formal bookkeeping device for
the occurrence of noise, we expand $Z_{\tau}^{(0)}$ and $Z_{\tau}^{(1)}$ in
powers of this parameter:
\begin{equation}
 Z_\tau^{(a)}
 =
 Z_{\tau,0}^{(a)}
 +\sqrt{\epsilon}\,Z_{\tau,1}^{(a)}
 +\epsilon Z_{\tau,2}^{(a)}
 +\cdots,
 \qquad\text{for }a=0,1.
 \label{eq:Zexpansion}
\end{equation}
In this equation, $Z_{\tau,0}^{(a)}$ is deterministic while other terms contain
randomness, and represent fluctuations.  

From the expansion of $Z_\tau^{(a)}$ of Eq. (\ref{eq:Zexpansion}), we can determine
approximations of $\mathbf{E}_{y}\big[q(Z_{\tau}^{(0)})\big]$ and 
$\mathbf{E}_{y}\big[q(Z_{\tau}^{(1)})\big]$ and use these in Eq. (\ref{Representation}) 
to obtain an approximation of $Q(\tau)$. This leads to the following.

\section*{Results}

The central result of this paper is an approximation for the population-level
allele-frequency statistic $Q(\tau)$ of Eq. (\ref{Q def}), that is valid in the
regime of strong selection, for $q(x)$ an arbitrary function\footnote{For our purposes, 
we shall assume that $q$ is smooth on the whole interval $[0,1]$, written $q \in C^\infty([0,1])$.}  
of $x$. The derivation
is given in Appendix \ref{small-noise expansion appendix}. The approximation
provides an answer to the question posed in the Introduction: how fixation
and loss combine to shape population-level statistics.

\subsubsection*{Details}

For the statistic $Q(\tau)$ we work to linear order in $\varepsilon$, discarding terms of order $\varepsilon^2$
and higher order.
We obtain $Q(\tau)\simeq Q_0(\tau)+\epsilon Q_1(\tau)$ and on reinstating the explicit form 
of  $\varepsilon$, namely $\varepsilon=1/R$, we obtain the
approximation\footnote{The approximation of $Q(\tau)$ in Eq.
(\ref{eq: Q firstorderapprox}) differs from an asymptotic expansion in $R$ - see
Section \ref{Appendix subsection expansion} of Appendix \ref{small-noise expansion appendix}.}
\begin{equation}
Q(\tau)\simeq Q_{0}(\tau)+\frac{Q_{1}(\tau)}{R}.  \label{eq: Q firstorderapprox}%
\end{equation}
% which involve $Z_{\tau,0}^{(a)}$, $M_{\tau}^{(a)}=??$ and $S_{\tau}^{(a)}=??$.
Using a prime to denote differentiation of a function with respect to its
argument, the quantities appearing in Eq. (\ref{eq: Q firstorderapprox}) are given by
\begin{align}
Q_{0}(\tau)  & =\langle q(Z_{\tau,0})\rangle_{\pi}\equiv\sum_{a=0}^{1}q\left(
Z_{\tau,0}^{(a)}\right)  \pi^{(a)}(y)\label{Q0 compact}\\[2ex]
Q_{1}(\tau)  & =\langle q^{\,\prime}(Z_{\tau,0})M_{\tau}\rangle_{\pi}+\frac
{1}{2}\langle q^{\,\prime\,\prime}(Z_{\tau,0})S_{\tau}\rangle_{\pi}\nonumber\\[1ex]
& \equiv\sum_{a=0}^{1}q^{\,\prime}(Z_{\tau,0}^{(a)})M_{\tau}^{(a)}\pi
^{(a)}(y)+\frac{1}{2}\sum_{a=0}^{1}q^{\,\prime\,\prime}(Z_{\tau,0}%
^{(a)})S_{\tau}^{(a)}\pi^{(a)}(y)\label{Q1 compact}%
\end{align}
where $Z_{\tau,0}^{(a)}$ represents two deterministic trajectories ($a=0,1$), while 
\begin{equation}
    M_{\tau}^{(a)}=\mathbf{E}_{y}\big[Z_{\tau,2}^{(a)}\big]\quad\text{and} \quad S_{\tau}^{(a)}=\mathbf{E}_{y}\Big[\big(Z_{\tau,1}^{(a)}\big)^{2}
\Big]     \label{fluctuation terms}
\end{equation}
are associated with fluctuations around each deterministic trajectory. 

The quantities $Z_{\tau,0}^{(a)}$, $M_{\tau}^{(a)}$, and $S_{\tau}^{(a)}$ are governed by the equations
\begin{equation}
\left.
\begin{array}
[c]{rcl}%
\dfrac{dZ_{\tau,0}^{(a)}}{d\tau} & = & f^{(a)}(Z_{\tau,0}^{(a)})\\[2ex]%
\dfrac{dM_{\tau}^{(a)}}{d\tau} & = & f^{(a)\,\prime}(Z_{\tau,0}^{(a)})\cdot
M_{\tau}^{(a)}+\frac{1}{2}f^{(a)\,\prime\,\prime}(Z_{\tau,0}^{(a)})\cdot
S_{\tau}^{(a)}\\[2ex]%
\dfrac{dS_{\tau}^{(a)}}{d\tau} & = & 2f^{(a)\,\prime}(Z_{\tau,0}^{(a)})\cdot
S_{\tau}^{(a)}+Z_{\tau,0}^{(a)}\big(1-Z_{\tau,0}^{(a)}\big)
\end{array}
\right\}  \label{ode system}%
\end{equation}
and subject to
\begin{equation}
Z_{0,0}^{(a)}=y,\qquad M_{0}^{(a)}=0,\qquad S_{0}^{(a)}=0.\label{init cond}%
\end{equation}
Equation (\ref{ode system}) is solved for $\tau\geq0$ under the initial
conditions of Eq. (\ref{init cond}). While a numerical solution is
usually required, a representation in terms of integrals is given in Appendix
\ref{integral representation appendix}.

\section*{Implications of the results}

There are some direct implications of the approximation of the statistic $Q(\tau)$ in Eq.
(\ref{eq: Q firstorderapprox}).

Following from Eq. (\ref{ode system}), the long-time limits of $Z_{\tau,0}^{(0)}$ and
$Z_{\tau,0}^{(1)}$ are $0$ and $1,$ respectively, while the long-time limits of
the fluctuation terms $M_{\tau}^{(a)}$ and $S_{\tau}^{(a)}$ (Eq. (\ref{fluctuation terms})) 
are both zero. This means the long-time limit of the approximation of $Q(\tau)$ 
is the exact
result
\begin{align}
\lim_{\tau\rightarrow\infty}Q(\tau)  & =\lim_{\tau\rightarrow\infty}%
\E_{y}[q(Y_{\tau})]\nonumber\\
& \nonumber\\
& =q\left(  0\right)  \pi^{(0)}(y)+q\left(  1\right)  \pi^{(1)}(y).  \label{lim Q exact}
\end{align}
This allows us to see some of the general problems with just using a single 
deterministic trajectory under strong selection.

For a beneficial allele, the deterministic trajectory
of Eq. (\ref{X deterministic}) tends to $1$ at long times.  If it were directly used to approximate
the allele-frequency mean, it would imply  $\lim_{\tau\rightarrow\infty}\E_y[Y_\tau]=1$.
This is generally incorrect, and particularly so when the probability of eventual loss is at all appreciable.
On using $q(x)=x$ in Eq. (\ref{lim Q exact}) the exact result is
\begin{equation}
 \lim_{\tau\rightarrow\infty}\E_y[Y_\tau] = \pi^{(1)}(y) \label{eq:meanlimit}
\end{equation}
and $\pi^{(1)}(y)$ can lie anywhere in the range $0$ to $1$, even under strong selection.

Similarly, on using $q(x)=x$ and $q(x)=x^2$ in Eq. (\ref{lim Q exact}), and
combining the results yields the exact result
\begin{equation}
 \lim_{\tau\rightarrow\infty}\Var_y(Y_\tau)
 =
 \pi^{(0)}(y)\pi^{(1)}(y).
 \label{eq:varlimit}
\end{equation}
By contrast, using the deterministic trajectory, along with fluctuations around it up to order $\varepsilon$,
to calculate the variance would yield a long-time value that vanishes.

The result in Eq. (\ref{eq:varlimit})
shows that the long-time form of the variance does not arise from 
fluctuations around a single deterministic
trajectory.  Rather, it is the outcome of the existence of
trajectories that achieve loss and fixation in different replicate populations.

The following numerical examples show that there can be transient behaviours, 
particularly of the variance, that illustrate the role of fluctuations at 
intermediate times.

\subsubsection*{Explicit strong-selection example}

We shall consider the regime of strong selection when the selection coefficient is positive ($s>0$).
For the purposes of illustration,  we shall use the same population size and selection coefficient as 
used in Figure 1. We thus have the census and effective population size coinciding and 
given by $N_e=3 \times 10^3$ and a selection coefficient of $s\simeq 1.67 \times 10^{-2}$ that 
lead to
\begin{equation*}
    R=2N_es=100.
\end{equation*}

Because we wish to compare results with the Wright-Fisher model, we chose initial allele
frequencies corresponding to an integer number of copies of the $A$ allele, 
namely $3$, $30$ and $300$ copies. These lead to initial frequencies of
\begin{equation*}
     y=5 \times 10^{-4},\quad 5 \times 10^{-3},\quad 5 \times 10^{-2},
\end{equation*}
respectively.  The strong-selection regime holds irrespective of the initial frequency, 
but the population-level importance of the two outcome classes is
very different for the three initial frequencies. 

For an initial frequency of $y=5 \times 10^{-4}$ 
the fixation probability is such that even though selection favours the $A$ allele, close
to $90\%$  of all trajectories achieve loss. For $y=5 \times 10^{-3}$ fixation occurs with a probability of
approximately $0.63$, and thus slightly exceeds loss, but for $y=5 \times 10^{-2}$ fixation 
is the strongly dominant outcome, with a probability of loss that is almost negligible (approximately 
$5 \times 10^{-5}$). The single deterministic trajectory in Eq. (\ref{X deterministic}), 
that is not conditioned in any way, but because $s>0$ proceeds to a frequency of $1$, 
does not represent any of these mixed outcomes correctly, but comes close to the last case. 

\subsubsection*{Mean allele frequency}

The mean is the simplest statistic for demonstrating the importance of
conditioning. From Eq. (\ref{eq: Q firstorderapprox}) with $q(x)=x$, we have
\begin{equation}
\mathbf{E}_{y}\left[  Y_{\tau}\right]  \simeq\langle Z_{\tau,0}\rangle_{\pi
}+\frac{\langle M_{\tau}\rangle_{\pi}}{R}.\label{eq:leadingapprox}%
\end{equation}
Since $\langle Z_{\tau,0}\rangle_{\pi}=Z_{\tau,0}^{(0)}\pi^{(0)}(y)+Z_{\tau
,0}^{(1)}\pi^{(1)}(y)$ we say 
\begin{equation}
\langle Z_{\tau,0}\rangle_{\pi} \text{ is the `weighted deterministic mean'}%
\end{equation}
and since $\langle M_{\tau}\rangle_{\pi}/R$ is associated with fluctuations
around the deterministic trajectories (see Eq. (\ref{fluctuation terms})), we say
\begin{equation}
\langle Z_{\tau,0}\rangle_{\pi}+\frac{\langle M_{\tau}\rangle_{\pi}}%
{R} \text{ is the `fluctuation-corrected mean'.}%
\end{equation}
In Figure 2 we plot both of these approximations of the mean as functions of the time, along with 
the exact mean allele frequency, which follows from the Wright-Fisher model.

\begin{figure}[H]
    \centering
    \includegraphics[width=1\linewidth]{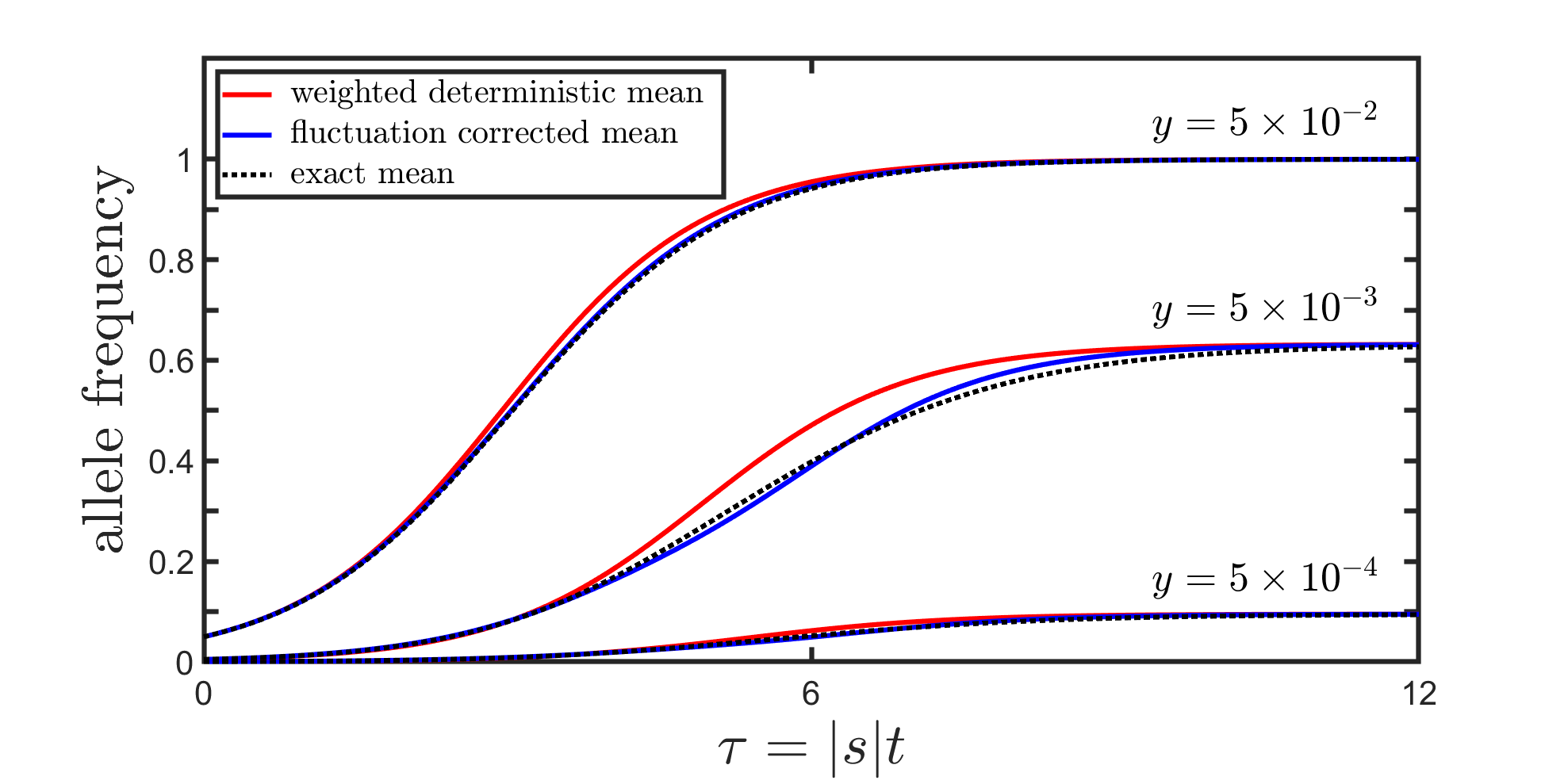}
    \caption{\textbf{The mean frequency as a function of time.} The figure
contains three well-separated sets of curves, one set for each of the initial
frequencies $y=5 \times 10^{-4}$, $y=5 \times 10^{-3}$, and $y=5 \times 10^{-2}$. \medskip\newline 
Each set, itself, contains three curves: (i) the `weighted deterministic mean' $\langle
Z_{\tau,0}\rangle_{\pi}$; (ii) the `fluctuation-corrected mean' 
$\langle Z_{\tau,0}\rangle_{\pi}+\langle M_{\tau
}\rangle_{\pi}/R$; and (iii) the exact Wright-Fisher result for the mean.\medskip\newline 
In the figure, the effective population size is equal to  the census population size. 
For all curves, the effective
population size is $N_{e}=3\times10^{3}$ and the selection coefficient is
$s\simeq 1.67 \times 10^{-2}$, corresponding to a scaled strength of selection of $R=100$. 
The three initial frequencies correspond to fixation probabilities of approximately
$0.1$, $0.63$, and $1$, respectively.}
    \label{fig:2}
\end{figure}

As follows from Eq. (\ref{lim Q exact}), the leading deterministic contribution, in the 
small-noise approximation ($\langle Z_{\tau,0}\rangle_{\pi}$) automatically  has the
correct long-time limit even before the contribution of within-trajectory fluctuations (the
$\langle M_{\tau}\rangle_{\pi}/R$ term) is included. 

Later we examine the accuracy of the approximation over a wide
range of parameters, but from this example it appears that the `weighted deterministic mean',
namely $\langle Z_{\tau,0}\rangle_{\pi}$, captures much of the behaviour of
the exact mean, and greater accuracy is obtained when the contribution from fluctuations is included. 

\bigskip

\subsubsection*{Variance of the allele frequency}

When the initial allele frequency is $y$, the variance of the allele frequency
at time $\tau$ is
\begin{equation}
\operatorname{Var}_{y}\left(  Y_{\tau}\right)  =\mathbf{E}_{y}\left[  \left(
Y_{\tau}\right)  ^{2}\right]  -\left(  \mathbf{E}_{y}\left[  Y_{\tau}\right]
\right)  ^{2}.
\end{equation}
The variance is the difference of two terms, which can be separately approximated from 
Eq. (\ref{eq: Q firstorderapprox}) using $q(x)=x^2$ and $q(x)=x$. 
Keeping terms to order $\varepsilon=1/R$ yields
\begin{equation}
\operatorname{Var}_{y}\left(  Y_{\tau}\right)  \simeq V_{0}+\frac{V_{1}}{R}
\label{Var Y approx}
\end{equation}
where 
\begin{equation}
V_{0}=\left\langle Z_{\tau,0}^{2}\right\rangle _{\pi}-\left\langle Z_{\tau
,0}\right\rangle _{\pi}^{2}\equiv{\left(  Z_{\tau,0}^{(0)}-Z_{\tau,0}
^{(1)}\right)  ^{2}\cdot\pi^{(0)}(y)\cdot\pi^{(1)}(y)} \label{V0}
\end{equation}
and
\begin{equation}
V_{1}=2\bigg(\left\langle Z_{\tau,0}M_{\tau}\right\rangle _{\pi}-\left\langle
Z_{\tau,0}\right\rangle _{\pi}\left\langle M_{\tau}\right\rangle _{\pi
}\bigg)+\left\langle S_{\tau}\right\rangle _{\pi}. \label{V1}
\end{equation}
These are both functions of $\tau$.

In Eq. (\ref{V0}), $V_{0}$ has been written in two different but
equivalent ways, one as a variance of $Z_{\tau,0}^{(a)}$,
namely ${\left\langle Z_{\tau,0}^{2}\right\rangle _{\pi}-\left\langle
Z_{\tau,0}\right\rangle _{\pi}^{2}}$, the other involving ${\left(  Z_{\tau
,0}^{(0)}-Z_{\tau,0}^{(1)}\right)  ^{2}}$. Both ways indicate that this
contribution to the variance arises because the two deterministic trajectories differ. 
In the long-time limit, $V_0$ tends to $\pi^{(0)}(y)\cdot\pi^{(1)}(y)$.
Furthermore, because the deterministic trajectories lie in the range $0$ to $1$,
 we have for all $\tau$ that $V_0 \le \pi^{(0)}(y)\cdot\pi^{(1)}(y)$.

 In Eq. (\ref{V1}), $V_1$ explicitly involves the fluctuation terms $M_{\tau}^{(a)}$ 
and $S_{\tau}^{(a)}$ of Eq. (\ref{fluctuation terms}). These both vanish in the long-time 
limit and hence so does $V_1$.

If we repeat the above calculation without conditioning, i.e., directly from Eq. (\ref{Y SDE}),
then the results are based on a single deterministic trajectory with corrections due to fluctuations.
This yields $V_0=0$, while $V_1$ arises from the single trajectory analogue of the fluctuation term 
$\left\langle S_{\tau}\right\rangle _{\pi}/R$. This term vanishes as $\tau \to \infty$, in contradiction to
Eq. (\ref{eq:varlimit}).

The variance thus provides a strong demonstration of why a conditioned `two-trajectory
representation' is needed.

Since $V_0$ originates from the contributions of two deterministic trajectories we say 
\begin{equation*}
V_0 \text{ is the `deterministic contribution to the variance'}%
\end{equation*}
and since $V_1/R$ is associated with fluctuations
around the deterministic trajectories (see Eq. (\ref{fluctuation terms})), we say
\begin{equation*}
V_0+V_1/R \text{ is  the `fluctuation-corrected variance'.}%
\end{equation*}
In Figure 3 we plot both of these approximations of the variance as functions of the time, 
along with 
the exact variance of the allele frequency, which follows from the Wright-Fisher model.

\begin{figure}[H]
    \centering
    \includegraphics[width=1\linewidth]{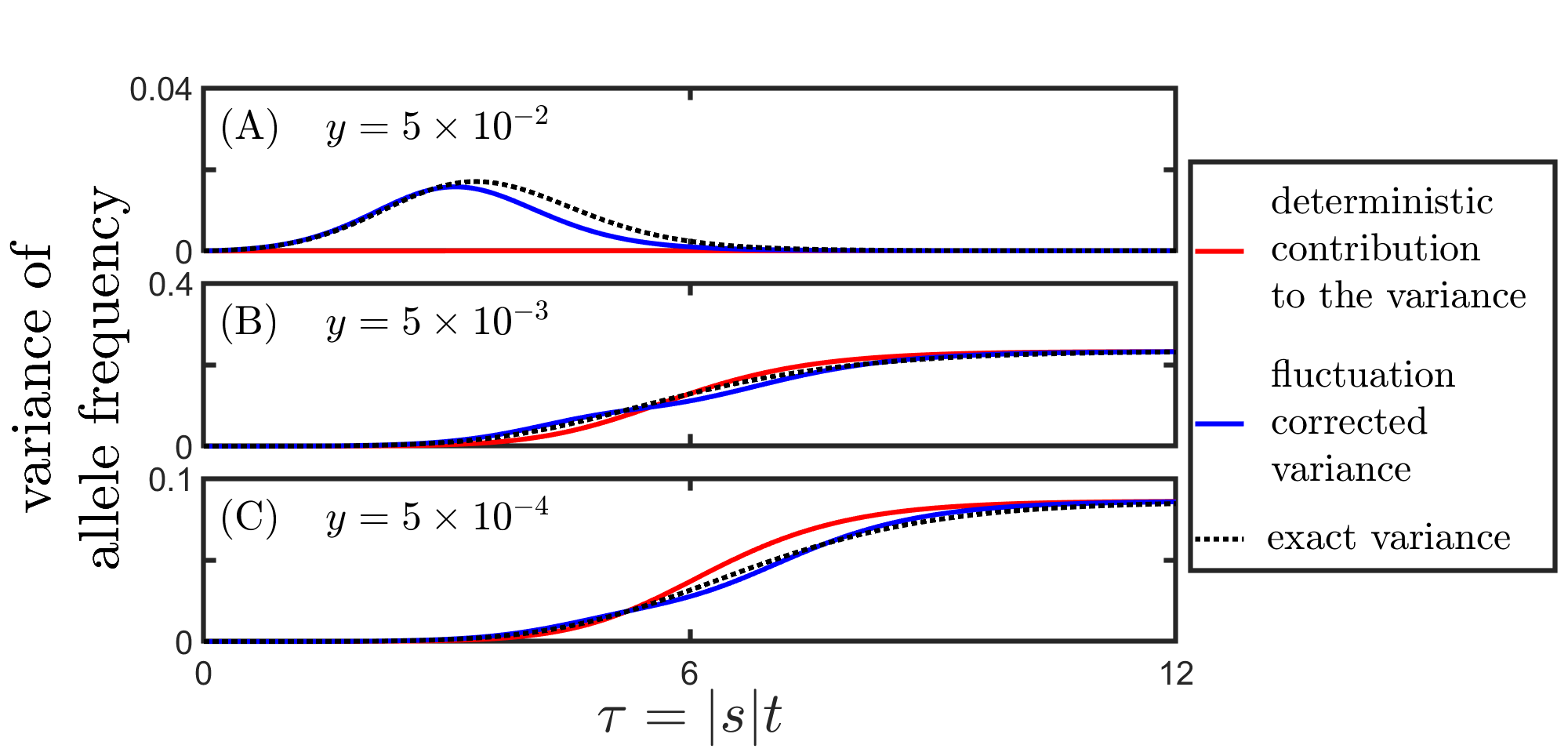}
    \caption{\textbf{Results for the variance of the allele frequency as
a function of time.} In the figure, the effective population size is equal to the census population size. Each panel shows three versions of the variance of the allele frequency as a function of time, $\tau$: 
(i) the deterministic contribution to the variance, $V_0$; (ii) the fluctuation-corrected variance, 
$V_0+V_1/R$; and (iii) the exact Wright–Fisher result.
\medskip\newline In all three panels, the effective population size is $N_e=3\times10^3$, and the selection coefficient is 
$s\simeq1.67\times10^{-2}$, corresponding to a scaled strength of selection of $R=100$. The panels differ only in the initial allele frequency, $y$.}
    \label{fig:3}
\end{figure}

It is evident from Figure 3 that the variance of the allele frequency can show a
range of different behaviours.

In Panel (A) of Figure 3 the deterministic contribution to the variance $V_{0}$  (red line), 
makes essentially no contribution to the variance over the time interval $0\leq\tau\lesssim 6$. 
This arises because $\pi^{(0)}(y)\cdot\pi^{(1)}(y) \ll 1$ when $R=100$ and $y=5 \times 10^{-2}$.
By contrast the fluctuation-corrected variance $V_{0}+V_{1}/R$ (blue line),
reasonably captures the variance over the same time interval. This means that 
essentially \textit{all} of the variance 
arises from the fluctuation term $V_{1}/R$ over this time interval. However, 
at large $\tau$ the fluctuation term $V_{1}/R$ tends to zero, and the deterministic contribution
then contributes all of the variance (which is very small).

By contrast, in Panels (B) and (C) of Figure 3, there is closer agreement between the 
deterministic contribution (red line) and the exact variance (dotted black line). However, the 
fluctuation-corrected variance (blue line) yields a yet closer agreement.

\section*{Numerical comparisons with the exact results}

\subsubsection*{Quantifying errors}

For the mean allele frequency and its variance, we compare the approximate
results that we have derived with the exact results. To make the
comparison we define%
\begin{equation}
\Delta= \text{error between an approximate and an exact statistic.}%\\
\end{equation}
Since all statistics considered in this work are time dependent, $\Delta$ is
calculated over the range of $\tau$ where the statistics exhibit variation.
Full details of $\Delta$ are given in Appendix \ref{error appendix}.
For practical purposes, we simply note that $\Delta \ge 0$, and values of $\Delta$ significantly below 
$100\%$ indicate a reasonable overall matching of exact and approximate forms, 
while large values indicate a mismatch.  

\subsubsection*{Errors on the mean and variance}

Taking the effective population size equal to the census size allows exact results
for the mean and variance, from the Wright-Fisher model, to be compared with
the approximate results given in Eqs. (\ref{eq:leadingapprox}) and
(\ref{Var Y approx}), respectively.

The Wright-Fisher model, with allele frequency $Y_{\tau}^{(WF)}$, yields
`numerically exact' statistics of the allele frequency at discrete
generations,  corresponding to $\tau$ equalling $0$, $|s|$, $2|s|$,\ldots\ (see
Eq. (\ref{tau Y sigma eps})). Since $|s|\ll1$ this is a very fine mesh and
such statistics can be continuously interpolated over $\tau$.

For the mean we define
\begin{align}
\Delta_{\text{det}}^{\text{(mean)}} &  =\text{error between the weighted
deterministic mean and the exact mean}\nonumber\\
& =\text{error between }\left\langle Z_{\tau,0}\right\rangle _{\pi
}\text{ and }\mathbf{E}_{y}\left[  Y_{\tau}^{(WF)}\right]  \nonumber\\[2ex]
\Delta_{\text{fluct}}^{\text{(mean)}} &  =\text{error between the fluctuation-corrected 
mean and the exact mean}\nonumber\\
& =\text{error between }\left\langle Z_{\tau,0}\right\rangle _{\pi
}+\frac{\left\langle M_{\tau}\right\rangle _{\pi}}{R}\text{ and }%
\mathbf{E}_{y}\left[  Y_{\tau}^{(WF)}\right]  \nonumber
\end{align}
and for the variance
\begin{align}
\Delta_{\text{det}}^{\text{(var)}} &  =\text{error between the
deterministic contribution and the exact variance}\nonumber\\
&  =\text{error between }V_{0}\text{ and }\operatorname{Var}_{y}\left(
Y_{\tau}^{(WF)}\right)  \nonumber\\[2ex]
\Delta_{\text{fluct}}^{\text{(var)}} &  =\text{error between the 
fluctuation-corrected variance and the exact variance}\nonumber\\
&  =\text{error between }V_{0}+\frac{V_{1}}{R}\text{ and }%
\operatorname{Var}_{y}\left(  Y_{\tau}^{(WF)}\right).  \nonumber
\end{align}

In Table 1 we give numerical results for $\Delta_{{\text{det}}}^{\text{(mean)}}$ and
$\Delta_{{\text{fluct}}}^{\text{(mean)}}$ for a set of parameter values.

\begin{table}[H]
\centering
\arrayrulecolor{black}
\begin{tabular}{|c|c|c|c|c|c|c|c|}
% \hline
% \textbf{Case} & \textbf{$s$} & \textbf{$R$} & \textbf{$y$} & \textbf{$R \times y$} & \textbf{$\pi^{(1)}(y)$} & \textbf{$\Delta_{\text{det}}^{\text{(mean)}}$} & \textbf{$\Delta_{\text{fluct}}^{\text{(mean)}}$} \\
% \hline
\hline
\vstrut\textbf{Case} & 
\vstrut\textbf{$s$} & 
\vstrut\textbf{$R$} & 
\vstrut\textbf{$y$} & \vstrut\textbf{$R \times y$} & 
\vstrut\textbf{$\pi^{(1)}(y)$} &  
\vstrut\textbf{$\Delta_{{\text{det}}}^{\text{(mean)}}\times 100$} & \vstrut\textbf{$\Delta_{{\text{fluct}}}^{\text{(mean)}}\times 100$} \\
\hline
\rowcolor{lightgray}
1 & -0.005 & 30 & 0.001 & 0.03 & 0.00 & 1.3 & 0.2 \\
\hline
\rowcolor{lightgray}
2 & -0.005 & 30 & 0.010 & 0.30 & 0.00 & 1.3 & 0.2 \\
\hline
\rowcolor{lightgray}
3 & -0.005 & 30 & 0.100 & 3.00 & 0.00 & 1.4 & 0.2 \\
\hline
\rowcolor{lightgray}
4 & -0.005 & 30 & 0.500 & 15.00 & 0.00 & 2.2 & 0.1 \\
\hline
5 & -0.010 & 60 & 0.001 & 0.06 & 0.00 & 0.4 & 0.5 \\
\hline
6 & -0.010 & 60 & 0.010 & 0.60 & 0.00 & 0.4 & 0.5 \\
\hline
7 & -0.010 & 60 & 0.100 & 6.00 & 0.00 & 0.4 & 0.4 \\
\hline
8 & -0.010 & 60 & 0.500 & 30.00 & 0.00 & 0.9 & 0.2 \\
\hline
\rowcolor{lightgray}
9 & -0.020 & 120 & 0.001 & 0.12 & 0.00 & 0.6 & 0.9 \\
\hline
\rowcolor{lightgray}
10 & -0.020 & 120 & 0.010 & 1.20 & 0.00 & 0.6 & 0.9 \\
\hline
\rowcolor{lightgray}
11 & -0.020 & 120 & 0.100 & 12.00 & 0.00 & 0.4 & 0.9 \\
\hline
\rowcolor{lightgray}
12 & -0.020 & 120 & 0.500 & 60.00 & 0.00 & 0.2 & 0.4 \\
\hline
13 & 0.005 & 30 & 0.001 & 0.03 & 0.06 & 9.7 & 3.1 \\
\hline
14 & 0.005 & 30 & 0.010 & 0.30 & 0.45 & 9.5 & 3.0 \\
\hline
15 & 0.005 & 30 & 0.100 & 3.00 & 1.00 & 4.1 & 0.8 \\
\hline
16 & 0.005 & 30 & 0.500 & 15.00 & 1.00 & 0.3 & 0.0 \\
\hline
\rowcolor{lightgray}
17 & 0.010 & 60 & 0.001 & 0.06 & 0.11 & 10.7 & 3.2 \\
\hline
\rowcolor{lightgray}
18 & 0.010 & 60 & 0.010 & 0.60 & 0.70 & 10.0 & 3.0 \\
\hline
\rowcolor{lightgray}
19 & 0.010 & 60 & 0.100 & 6.00 & 1.00 & 2.1 & 0.3 \\
\hline
\rowcolor{lightgray}
20 & 0.010 & 60 & 0.500 & 30.00 & 1.00 & 0.1 & 0.0 \\
\hline
21 & 0.020 & 120 & 0.001 & 0.12 & 0.21 & 11.8 & 3.4 \\
\hline
22 & 0.020 & 120 & 0.010 & 1.20 & 0.91 & 9.7 & 2.4 \\
\hline
23 & 0.020 & 120 & 0.100 & 12.00 & 1.00 & 1.1 & 0.2 \\
\hline
24 & 0.020 & 120 & 0.500 & 60.00 & 1.00 & 0.0 & 0.1 \\
\hline
\end{tabular}
\caption{\textbf{Errors in approximations of the mean allele frequency.}  When the mean allele frequency is approximated by
just the deterministic term, $\left\langle Z_{\tau,0}\right\rangle
_{\pi}$, the error is $\Delta_{{\text{det}}}^{\text{(mean)}}$. When the approximation includes
deterministic and fluctuation contributions, i.e., $\left\langle Z_{\tau,0}\right\rangle _{\pi
}+\left\langle M_{\tau}\right\rangle _{\pi}/R$, the error is $\Delta_{{\text{fluct}}}^{\text{(mean)}}$.
 For the table, the census and effective population sizes were set equal, with the value 
 $N_{e}=3 \times10^{3}$.}
\end{table}

Table 1 gives an indication of the overall accuracy of using $\left\langle
Z_{\tau,0}\right\rangle _{\pi}$ or $\left\langle Z_{\tau,0}\right\rangle
_{\pi}+\left\langle M_{\tau}\right\rangle _{\pi}/R$ in place of the exact
result for the mean. It can be seen that $\Delta_{{\text{fluct}}}^{\text{(mean)}}$ is
typically smaller than $\Delta_{{\text{det}}}^{\text{(mean)}}$.

In Table 2 we give numerical results for $\Delta_{{\text{det}}}^{\text{(var)}}$ and
$\Delta_{{\text{fluct}}}^{\text{(var)}}$ for a set of parameter values.

%TABLE 2 HERE

\begin{table}[H]
\centering
\arrayrulecolor{black}
\begin{tabular}{|c|c|c|c|c|c|c|c|}
\hline
\vstrut\textbf{Case} &
\vstrut\textbf{$s$} & 
\vstrut\textbf{$R$} & 
\vstrut\textbf{$y$} & 
\vstrut\textbf{$R \times y$} & 
\vstrut\textbf{$\pi^{(1)}(y)$} & 
\vstrut\textbf{$\Delta_{{\text{det}}}^{\text{(var)}}\times 100$} & \vstrut\textbf{$\Delta_{{\text{fluct}}}^{\text{(var)}}\times 100$} \\
\hline
\rowcolor{lightgray}
1 & -0.005 & 30 & 0.001 & 0.03 & 0.00 & 100.0 & 3.0 \\
\hline
\rowcolor{lightgray}
2 & -0.005 & 30 & 0.010 & 0.30 & 0.00 & 100.0 & 3.0 \\
\hline
\rowcolor{lightgray}
3 & -0.005 & 30 & 0.100 & 3.00 & 0.00 & 100.0 & 3.3 \\
\hline
\rowcolor{lightgray}
4 & -0.005 & 30 & 0.500 & 15.00 & 0.00 & 100.0 & 5.4 \\
\hline
5 & -0.010 & 60 & 0.001 & 0.06 & 0.00 & 100.0 & 1.1 \\
\hline
6 & -0.010 & 60 & 0.010 & 0.60 & 0.00 & 100.0 & 1.1 \\
\hline
7 & -0.010 & 60 & 0.100 & 6.00 & 0.00 & 100.0 & 1.3 \\
\hline
8 & -0.010 & 60 & 0.500 & 30.00 & 0.00 & 100.0 & 2.6 \\
\hline
\rowcolor{lightgray}
9 & -0.020 & 120 & 0.001 & 0.12 & 0.00 & 100.0 & 0.3 \\
\hline
\rowcolor{lightgray}
10 & -0.020 & 120 & 0.010 & 1.20 & 0.00 & 100.0 & 0.3 \\
\hline
\rowcolor{lightgray}
11 & -0.020 & 120 & 0.100 & 12.00 & 0.00 & 100.0 & 0.2 \\
\hline
\rowcolor{lightgray}
12 & -0.020 & 120 & 0.500 & 60.00 & 0.00 & 100.0 & 1.1 \\
\hline
13 & 0.005 & 30 & 0.001 & 0.03 & 0.06 & 11.9 & 4.7 \\
\hline
14 & 0.005 & 30 & 0.010 & 0.30 & 0.45 & 9.7 & 4.7 \\
\hline
15 & 0.005 & 30 & 0.100 & 3.00 & 1.00 & 90.4 & 26.4 \\
\hline
16 & 0.005 & 30 & 0.500 & 15.00 & 1.00 & 100.0 & 5.4 \\
\hline
\rowcolor{lightgray}
17 & 0.010 & 60 & 0.001 & 0.06 & 0.11 & 12.6 & 4.9 \\
\hline
\rowcolor{lightgray}
18 & 0.010 & 60 & 0.010 & 0.60 & 0.70 & 9.1 & 6.5 \\
\hline
\rowcolor{lightgray}
19 & 0.010 & 60 & 0.100 & 6.00 & 1.00 & 100.0 & 14.8 \\
\hline
\rowcolor{lightgray}
20 & 0.010 & 60 & 0.500 & 30.00 & 1.00 & 100.0 & 2.6 \\
\hline
21 & 0.020 & 120 & 0.001 & 0.12 & 0.21 & 12.9 & 5.1 \\
\hline
22 & 0.020 & 120 & 0.010 & 1.20 & 0.91 & 49.0 & 19.6 \\
\hline
23 & 0.020 & 120 & 0.100 & 12.00 & 1.00 & 100.0 & 7.9 \\
\hline
24 & 0.020 & 120 & 0.500 & 60.00 & 1.00 & 100.0 & 1.1 \\
\hline
\end{tabular}
\caption{\textbf{Errors in approximations of the variance.}  When the variance of
the allele frequency is approximated by just the deterministic term, $V_0$, the error
is $\Delta_{{\text{det}}}^{\text{(var)}}$. When the approximation includes
deterministic and fluctuation contributions i.e., $V_0+V_1/R$, the error
is $\Delta_{{\text{fluct}}}^{\text{(var)}}$. For the table, the census and effective population 
sizes were set equal, with the value $N_{e}=3 \times10^{3}$.}
\end{table}

\bigskip

Table 2 gives an indication of the overall accuracy of using $V_{0}$ or
$V_{0}+V_{1}/R$ in place of the exact result for the variance. It can be seen
that in a number of cases, the purely deterministic contribution $V_{0}$ `fails' as
indicated by very large values of $\Delta_{\text{det}}^{\text{(var)}}$ in the
table. This arises from a behaviour similar to that seen in Panel A of Figure
3, where  $V_0$ has a very small value and hence makes a very small contribution. 
However, in such cases the fluctuation contribution, $V_{1}/R$, which is included in the 
fluctuation-corrected variance, is sufficient to greatly reduce the error, 
as seen in the corresponding values of $\Delta_{\text{fluct}}^{\text{(var)}}$. Additionally, in the cases where 
$\Delta_{\text{det}}^{\text{(var)}}$ is not large, the contribution of the fluctuation 
term still has the effect of improving accuracy, as seen in the corresponding values of $\Delta_{\text{fluct}}^{\text{(var)}}$.

\section*{Discussion}

This paper began with the question: when selection is strong relative to
genetic drift, how do fixation and loss combine to shape population-level
statistics of allele frequency? The answer developed here is that while
strong selection makes some individual trajectories comparatively
deterministic, the population-level distribution of allele frequencies does
not reflect such determinism. Working on timescales where mutation can be
neglected, trajectories fall into two absorbing classes: those that
eventually fix and those that are eventually lost. The population-level
distribution of allele frequencies and associated statistics, which follow
from an average across replicate populations, are shaped by: (i) the level
of determinism within each class of trajectory and (ii) the probabilities of
the different long-time outcomes.

For the determination of statistics, a fundamental way to deal with such
issues is to \textit{condition} on both outcomes, and then weight the
conditioned results with the respective probabilities of the two outcomes.
This approach directly forces both classes of trajectory into the analysis.

The reason conditioning is necessary, rather than merely convenient, is that
a single deterministic trajectory cannot represent two absorbing outcomes.
Without conditioning, the leading-order result can be wrong by an amount that
does not shrink as the noise gets smaller. Within each conditioned class,
however, there is only one outcome, so a deterministic trajectory is the
right starting point and the fluctuations are genuinely small corrections. The
difficulty is not the diffusion approximation, which describes the fate of
the allele well \citep{Shafiey2017}, but the existence of two possible
outcomes.

Under strong selection, it is not surprising that a trajectory drawn from the
class that eventually achieves fixation has the form of a deterministic
trajectory plus small corrections due to fluctuations. It is perhaps surprising
that a trajectory drawn from the class that eventually achieves loss also
has the form of a deterministic trajectory plus small corrections due to fluctuations. 
This arises because conditioning is equivalent to an underlying
unconditioned dynamics that has ``fictitious selection''
\citep{Ewens1973,ZhaoLascouxOverall}. Fictitious selection ensures the appropriate
outcome occurs, by preventing the other outcome. The selective force in the
equivalent unconditioned dynamics (which includes the fictitious selection)
is strong compared with drift. This allows use of a small-noise
approximation, which recovers the correct long-time values of statistics and
captures non-trivial transient behaviours.

The above results make it clear that there is a distinction between the
level of determinism within an outcome, and the heterogeneity of
outcomes. Strong positive selection means that once an allele is at
appreciable frequency, deterministic directional change dominates drift. It
does not mean that an initially rare beneficial allele necessarily
establishes, and it does not mean that fixation is assured. Selection,
establishment, and fixation are three distinct concepts that the framework
developed here separates naturally. For a beneficial mutation, the
loss-conditioned trajectory represents early loss, the fixation-conditioned
trajectory represents successful establishment and subsequent increase, and
the probabilities of loss and fixation connect the two trajectory classes to
population-level statistics.

\subsubsection*{Selection strength, initial frequency, and mutation}

The condition $R=2N_{e}|s|\gg1$ defines strong selection relative to genetic
drift. It does not require large selection coefficients. Table~\ref%
{tab:regimes} gives parameter combinations spanning several biological
regimes.

\begin{table}[H]
\centering
{\small \setlength{\tabcolsep}{6pt} \renewcommand{\arraystretch}{1.3} 
\begin{tabular}{|c|c|c|c|l|}
\hline
\multicolumn{1}{|c|}{\textbf{$N_{e}$}} & \multicolumn{1}{c|}{\textbf{$|s|$}}
& \multicolumn{1}{c|}{\textbf{$R = 2N_{e}|s|$}} & \multicolumn{1}{c|}{%
\textbf{timescale $\sim1/|s|$}} & \multicolumn{1}{c|}{\textbf{Biological
setting}} \\ \hline
$10^{3}$ & $10^{-2}$ & $20$ & $100$ & small experimental/lab population \\ 
\hline
$10^{4}$ & $10^{-3}$ & $20$ & $1000$ & moderate natural population \\ \hline
$10^{5}$ & $10^{-3}$ & $200$ & $1000$ & large natural population \\ \hline
$10^{6}$ & $10^{-3}$ & $2000$ & $1000$ & large cellular population \\ \hline
\end{tabular}
}
\caption{\textbf{Parameter regimes for strong selection relative to drift.}
The composite parameter $R = 2N_e|s|$ measures the strength of selection 
relative to drift,
and the characteristic timescale of allele-frequency changes is $\sim 1/|s|$
generations. Here $N_e$ is the effective population size and $s$ is the
additive selection coefficient. The values are illustrative rather than
universal.}
\label{tab:regimes}
\end{table}

While strong selection is expressed in terms of the size of $R$, the
probabilities of the two classes of trajectory are controlled not only by $R$
but also by the scaled initial frequency $Ry$. When $Ry$ is small, the
fixation probability, $\pi ^{(1)}(y)$, can be substantially less than one
even under strong selection. For example, with $R=100$ and $y=5\times 10^{-4}
$, the fixation probability is approximately $0.1$, hence close to $90\%$ of
trajectories achieve loss. With the same $R$ and $y=5\times 10^{-2}$,
fixation is almost certain. The intermediate case, $y=5\times 10^{-3}$,
gives a fixation probability of approximately $0.63$, so both outcome
classes occur at broadly comparable levels. The three regimes, of
loss-dominated, mixed, and fixation-dominated, correspond to qualitatively
different biological situations, but the framework developed here treats all
three within a single mathematical structure.

The probability that a particular trajectory-class occurs may not directly measure
its contribution to a statistic. While for some statistics such as the mean
allele frequency, the probabilities of fixation and loss occur linearly, as
weightings of conditioned statistics, and in this case these probabilities
are a measure of a class's contribution. For other statistics, such as the 
variance, the leading-order result is proportional to $\pi^{(0)}(y)\pi^{(1)}(y)$. 
The dependence on this product means that both classes of trajectory contribute 
jointly, and the statistic cannot be attributed to (or dominated by) a single class.

In this work we neglect mutation. Whether this is a reasonable approximation depends 
on how mutation affects the biallelic dynamics and on the timescale considered.

Recurrent mutation between the focal and alternative alleles can make fixation and 
loss non-absorbing, even though the locus remains biallelic.

Mutation to a third allele, by contrast, takes the locus outside the biallelic model. 
The probability that a new mutant establishes depends on its fitness relative to the 
resident population and, where relevant, on dominance, genetic background, and population size.
Establishment cannot in general be determined from the selection coefficient and frequency 
of the focal allele alone.

For either type of mutation, the decomposition used in this work, which conditions on eventual 
fixation or loss, would not apply in its present form. Assessing the importance of mutation 
requires information about mutation rates and targets, the fitness effects of new alleles, 
and the timescale of interest. The neglect of mutation is therefore a modelling assumption 
whose validity depends on the biological system and the question being addressed.

\subsubsection*{Connection to population genomics}

Population genomics provides allele-frequency and haplotype data across
populations and through time. Genome scans for recent positive selection
commonly use signatures such as reduced local diversity, distortions of the
site-frequency spectrum, long-range haplotype structure, linkage
disequilibrium, and allele-frequency changes through time %
\citep{Nielsen,Sabeti,Messer,Pavlidis2017}. The present model
addresses a narrower but important component of this problem: the
time-dependent distribution of the selected allele itself. It does not model
linked neutral sites, recombination, or the full genomic footprint of a
sweep. But it does provide a trajectory-level description that can be
coupled to more complex models.

The practical relevance is most direct when a selected allele is observed
before its ultimate fate is known. In a population-genomic time series, an
allele that is increasing rapidly may be on a trajectory destined for
fixation, but the ensemble of populations or replicate experiments can also
contain trajectories that are ultimately lost. Treating all such
observations as noisy realisations of a single deterministic logistic path
(see Eq. (2) and the deterministic trajectory plotted in blue in Figure 1),
is a natural approach under the assumption that strong selection implies a
single deterministic trajectory. This can severely misrepresent the
distribution of frequencies, particularly when the allele begins at low
frequency and the probability of loss is appreciable. In that regime, the
population-level mean frequency at long times is not $1$ but the fixation
probability $\pi ^{(1)}(y)$, and the population-level variance is not zero
but $\pi ^{(0)}(y)\pi ^{(1)}(y)$. These long-time results transcend the
diffusion approximation and follow directly from a two-outcome decomposition.

The framework may also be relevant to the interpretation of partial sweeps.
A partial sweep, where an allele has risen to intermediate frequency but has
not yet fixed, is not itself evidence that the allele will ultimately fix.
The present model makes the distinction between current frequency and the
eventual outcome explicit, and it provides a forward model in which both
outcomes are included rather than assumed away.

\subsubsection*{Applications and future directions}

Three application areas motivate future developments.

\paragraph{Experimental evolution.}

Replicated experimental populations sampled over time allow the distribution
of allele frequencies across replicates to be measured directly. The
fixation--loss representation provides a natural viewpoint for disentangling
two statistically distinct sources of variation: differences between
replicates in eventual absorbing outcome, and fluctuations around the
conditional trajectory. In experimental microbial systems, where replicate
populations can be followed for many generations, this distinction is likely
to be important for interpreting the spread of beneficial alleles.

\paragraph{Longitudinal natural populations.}

Temporal allele-frequency data, including ancient DNA and other time-series
sampling schemes, contain information about selection strength, but
inference is complicated by drift and by uncertainty about the future fate.
The conditional formulation provides a forward model for such data under
strong selection. Such a model could potentially be used as a component of
likelihood-based or simulation-based inference for selection strength.

\paragraph{Rapid adaptation.}

Strong selection occurs in microbial and other rapidly evolving populations,
where repeated sampling reveals transient allele-frequency trajectories.
Distinguishing deterministic change within the class with a given fate, from
stochastic variation in establishment is likely to be important. The same
logic applies to conservation genomics, where predicting the fate of
beneficial alleles in small populations under strong selection is of
practical interest.

\subsubsection*{Scope and limitations}

The present model has several deliberate limitations, many of which were
made to make the analysis feasible.

\begin{enumerate}
\item \textbf{Single biallelic locus.} The two-class picture of fixation and
loss is specific to a biallelic locus. With more than two alleles, only
fixation of one allele uniquely determines the state of all others while
loss does not. A treatment for more than two alleles would have to represent
a basic statistic as a sum over fixation-conditioned expectations weighted
by the corresponding fixation probabilities of the different alleles. This
is a natural extension but is not pursued here.

\item \textbf{Additive selection with constant $s$.} Selection is assumed
additive and represented by a constant coefficient. Dominance, epistasis,
and fluctuating selection are not included. The framework could be extended
to include dominance and epistasis via modified selection functions.
Fluctuating selection may require additional considerations.

\item \textbf{No mutation.} Mutation is neglected over the timescales of
interest. This is what gives the dynamics two absorbing states. If mutation
is included, the picture changes qualitatively: mutation, selection, and
drift co-occur, and the two-absorbing-state decomposition is no longer
applicable in its present form. A different analysis would be required.

\item \textbf{Constant $N_{e}$.} Population size is represented by a
constant effective size. Changing demographic history, population structure,
and migration are not included. These would introduce additional sources of
heterogeneity that could compete with or compound the between-outcome
heterogeneity studied here.

\item \textbf{No recombination or linkage.} The model treats a single locus
and does not represent linked neutral sites. Coupling the
outcome-conditioned selected-allele frequencies to a model of linked neutral
ancestry would provide a direct bridge to genomic sweep statistics, for
example, by asking how conditioning on fixation versus loss of the selected
allele changes the expected distribution of linked diversity and haplotype
lengths.
\end{enumerate}

\subsubsection*{Concluding remarks}

By conditioning separately on the two absorbing outcomes, the present
approach expresses statistics in terms of two deterministic trajectories and
corrections from fluctuations. The computational cost is extremely modest:
approximate results for population-level statistics require only the
solution of a small set of differential equations. The present framework
provides a tractable starting point for further analyses and may have
extensions to experimental evolution, longitudinal population-genomic
studies, and models of selective sweeps.

\newpage 

\begin{center}
{\Large {\textbf{APPENDICES}}}
\end{center}

\appendix
\renewcommand{\thesection}{\arabic{section}} % Makes appendix labels 1,2,.. instead of A, B, ..
\numberwithin{equation}{section}

The appendices below provide the full mathematical derivations underlying the results in the main text. 
Their order follows the logic of the main text: time rescaling, outcome of conditioning, small-noise expansion, integral representations, and error quantification.

\section[Time transformation of the dynamical equation of X(t)]{Time transformation of the dynamical equation of $X_{t}$}

\label{transformation SDE appendix}

In this appendix we produce a time-transformed version of the stochastic
differential equation satisfied by the random frequency $X_{t}$, given by Eq.
(\ref{X SDE}). While we can derive the time-transformed equation from
Eq. (\ref{X SDE}), there are technical points that have to be properly
handled, and we proceed in a simpler but equivalent approach.

\subsection{Background}

For the purposes of this appendix, we consider the following It\^{o}
stochastic differential equation for $X_{t}$:
\begin{equation}
dX_{t}=a(X_{t})\,dt+b(X_{t})dW_{t} \label{general SDE}
\end{equation}
where $W_{t}$ is a standard Wiener process \citep{Tuckwell}. We take the
initial time to be $0$ and the initial value of $X_{t}$ to be $y$:
\begin{equation}
X_{0}=y.
\end{equation}

We write the distribution (probability density) of $X_{t}$, when evaluated at
$x$, as $K(x,t|y,0)$. The equation that $K(x,t|y,0)$ obeys (the forward
Kolmogorov/Fokker-Planck/diffusion equation) directly follows from Eq.
(\ref{general SDE}) \citep{Tuckwell} and is given by
\begin{equation}
-\frac{\partial}{\partial t}K(x,t\,|\,y,0)=-\frac{1}{2}\frac{\partial^{2}
}{\partial x^{2}}\left[  \left[  b(x)\right]  ^{2}K(x,t\,|\,y,0)\right]
+\frac{\partial}{\partial x}\left[  a(x)K(x,t\,|\,y,0)\right]  .
\label{General K eq}
\end{equation}
This equation is a description of the problem that is fully equivalent\footnote{We leave implicit
that $K(x,0\,|\,y,0)=\delta(x-y)$, a Dirac delta function, and that probability is conserved \citep{McKane_Waxman}.}
to Eq.
(\ref{general SDE}).

\subsection{Time change}

With $c$ a positive constant, we introduce the rescaled time
\begin{equation}
\tau=c\times t
\end{equation}
and define
\[
\widetilde{K}(x,\tau\,|\,y,0)=K(x,t\,|\,y,0)\equiv K\left(  x,\frac{\tau}
{c}\,|\,y,0\right)  .
\]
We apply the chain rule to $t$ derivatives in Eq. (\ref{General K eq}), and in
terms of $\widetilde{K}(x,\tau\,|\,y,0)$ we have
\begin{equation}
-\frac{\partial}{\partial\tau}\widetilde{K}(x,\tau\,|\,y,0)=-\frac{1}{2}
\frac{\partial^{2}}{\partial x^{2}}\left[  \left(  \frac{b(x)}{\sqrt{c}
}\right)  ^{2}\widetilde{K}(x,\tau\,|\,y,0)\right]  +\frac{\partial}{\partial
x}\left[  \frac{a(x)}{c}\widetilde{K}(x,\tau\,|\,y,0)\right]  .
\label{K transformed}
\end{equation}
From the relation between the equivalent descriptions in Eqs.
(\ref{general SDE}) and (\ref{General K eq}) we infer that Eq.
(\ref{K transformed}) is equivalent to a stochastic differential equation for
\begin{equation}
Y_{\tau}=X_{t}\equiv X_{\tau/c}
\end{equation}
that is given by
\begin{equation}
dY_{\tau}=\frac{a(Y_{\tau})}{c}\,d\tau+\frac{b(Y_{\tau})}{\sqrt{c}}dW_{\tau}.
\end{equation}
This is the result of the time transformation on Eq. (\ref{general SDE}).

For Eq. (\ref{X SDE}) of the main text we have $a(x)=sx(1-x)$,
$b(x)=\sqrt{\frac{x(1-x)}{2N_{e}}}$ and $c=|s|$ and the transformed version of
Eq. (\ref{X SDE}) is
\begin{equation}
dY_{\tau}=\frac{s}{|s|}\,Y_{\tau}\left(  1-Y_{\tau}\right)  \,d\tau
+\sqrt{\frac{Y_{\tau}\left(  1-Y_{\tau}\right)  }{2N_{e}|s|}}\,dW_{\tau}.
\end{equation}
This is equivalent to Eq. (\ref{Y SDE}) of the main text.

%%%%%%%%%%%%%%%%%%%%%%%%%%%%%%%%%%%%%%%%%%%%%%%%%%%%%%%%%%%%%%%%%%%%%%%%%%%%%

\section{Properties of the modified random frequencies}

\label{modified dynamics appendix}

In this appendix, we establish properties of two random frequencies that can
serve as substitutes for $Y_{\tau}$ when conditioned on either fixation or loss.

We begin with the equation that $Y_{\tau}$ obeys (Eq. (\ref{Y SDE}) of the
main text), written in the form
\begin{equation}
dY_{\tau}=\sigma\,Y_{\tau}\left(  1-Y_{\tau}\right)  \,d\tau+\sqrt
{\frac{Y_{\tau}\left(  1-Y_{\tau}\right)  }{R}}\,dW_{\tau}
.\label{appendix Y SDE}
\end{equation}
Let $K(x,\tau\,|\,y,0)$ denote the probability density of $Y_{\tau}$ when
evaluated at $x$. This obeys
\begin{align}
-\frac{\partial}{\partial\tau}K(x,\tau\,|\,y,0) &  =-\frac{1}{2R}
\frac{\partial^{2}}{\partial x^{2}}\left[  x(1-x)K(x,\tau\,|\,y,0)\right]
\nonumber\\
& \nonumber\\
&  \quad+\frac{\partial}{\partial x}\left[  \sigma x(1-x)K(x,\tau
\,|\,y,0)\right]  .\label{K pde}
\end{align}

We write the corresponding probability density, when $Y_{\tau}$ is conditioned
on the ultimate occurrence of fixation, as $K^{(1)}(x,\tau\,|\,y,0)$. With
$\pi^{(1)}(y)=\left(  1-e^{-2\sigma Ry}\right)  /\left(  1-e^{-2\sigma
R}\right)  $ the probability of fixation when $Y_{0}=y$ we have
\begin{equation}
K^{(1)}(x,\tau\,|\,y,0)=\pi^{(1)}(x)K(x,\tau\,|\,y,0)/\pi^{(1)}(y)\label{K1}
\end{equation}
\citep{ZhaoLascouxOverall,Ewens1973}.

The equation obeyed by $K^{(1)}(x,\tau\,|\,y,0)$ follows from Eq.~(\ref{K pde}
) by substituting the `inverse' of Eq.~(\ref{K1}), namely $K(x,\tau
\,|\,y,0)=\pi^{(1)}(y)K^{(1)}(x,\tau\,|\,y,0)/\pi^{(1)}(x)$, into
Eq.~(\ref{K pde}). Some algebra, involving the fact that $\pi^{(1)}(y)$
obeys the time-independent backward equation
\begin{equation}
-\frac{y(1-y)}{2R}\frac{d^{2}}{dy^{2}}\pi^{(1)}(y)-\sigma y(1-y)\frac{d}
{dy}\pi^{(1)}(y)=0
\end{equation}
leads to $K^{(1)}(x,\tau\,|\,y,0)$ obeying
\begin{align}
-\frac{\partial}{\partial\tau}K^{(1)}(x,\tau\,|\,y,0) &  =-\frac{1}{2R}
\frac{\partial^{2}}{\partial x^{2}}\left[  x(1-x)K^{(1)}(x,\tau
\,|\,y,0)\right]  \nonumber\\
& \nonumber\\
&  \quad+\frac{\partial}{\partial x}\left[  \coth(Rx)\,x(1-x)K^{(1)}
(x,\tau\,|\,y,0)\right]  .\label{K1 pde}
\end{align}
Equation (\ref{K1 pde}) can be seen to follow from Eq.~(\ref{K pde}) by the
replacement of $\sigma$ by $\coth(Rx)$. 

Comparing Eqs. (\ref{appendix Y SDE}) and (\ref{K pde}) shows that
Eq.~(\ref{K1 pde}) governs the distribution of a random frequency, denoted by
$Z_{\tau}^{(1)}$, which obeys $dZ_{\tau}^{(1)}=\coth(RZ_{\tau}^{(1)})\,Z_{\tau}^{(1)}(1-Z_{\tau}
^{(1)})\,d\tau+\sqrt{Z_{\tau}^{(1)}(1-Z_{\tau}^{(1)})/R}\,dW_{\tau}$
and with
\begin{equation}
\varepsilon=1/R
\end{equation}
we can write the equation for $Z_{\tau}^{(1)}$ as
\begin{equation}
dZ_{\tau}^{(1)}=\coth(RZ_{\tau}^{(1)})\,Z_{\tau}^{(1)}(1-Z_{\tau}
^{(1)})\,d\tau+\sqrt{\varepsilon}\cdot\sqrt{Z_{\tau}^{(1)}(1-Z_{\tau}^{(1)}
)}\,dW_{\tau}.\label{appendix Z1 SDE}
\end{equation}
Furthermore, the distribution of $Z_{\tau}^{(1)}$, \textit{without any
conditioning}, coincides with the conditioned distribution of $Y_{\tau}$,
namely $K^{(1)}(x,\tau\,|\,y,0)$. In other words averaging over a function of
$Z_{\tau}^{(1)}$, without any conditioning, is equivalent to averaging over
the same function of $Y_{\tau}$, when conditioned upon fixation. This means,
with $fix$ shorthand for `the ultimate occurrence of fixation', that for an
integrable function $q(\cdot)$
\begin{equation}
\mathbf{E}_{y}\left[  q(Y_{\tau})\,|\,fix\right]  =\mathbf{E}_{y}\left[
q\left(  Z_{\tau}^{(1)}\right)  \right]  .\label{mod E 1}
\end{equation}
We describe $Z_{\tau}^{(1)}$ as a \textit{modified version} of $Y_{\tau}$.

In a similar way, the probability density, when $Y_{\tau}$ is conditioned on
the ultimate occurrence of loss, written $K^{(0)}(x,\tau\,|\,y,0)$, is given
by
\begin{equation}
K^{(0)}(x,\tau\,|\,y,0)=\pi^{(0)}(x)K(x,\tau\,|\,y,0)/\pi^{(0)}(y)\label{K0}
\end{equation}
where $\pi^{(0)}(x)=1-\pi^{(1)}(x)$ is the probability of loss. Repeating the
above steps for $K^{(0)}(x,\tau\,|\,y,0)$ leads to
\begin{align}
-\frac{\partial}{\partial\tau}K^{(0)}(x,\tau\,|\,y,0) &  =-\frac{1}{2R}
\frac{\partial^{2}}{\partial x^{2}}\left[  x(1-x)K^{(0)}(x,\tau
\,|\,y,0)\right]  \nonumber\\
& \nonumber\\
&  \quad-\frac{\partial}{\partial x}\left[  \coth(R(1-x))\,x(1-x)K^{(0)}
(x,\tau\,|\,y,0)\right]  .\label{K0 pde}
\end{align}
This tells us that there is a modified version of $Y_{\tau}$, denoted 
$Z_{\tau}^{(0)}$, that obeys
\begin{equation}
dZ_{\tau}^{(0)}=-\coth\left(  R\left(  1-Z_{\tau}^{(0)}\right)  \right)
\,Z_{\tau}^{(0)}(1-Z_{\tau}^{(0)})\,d\tau+\sqrt{\varepsilon}\cdot\sqrt
{Z_{\tau}^{(0)}(1-Z_{\tau}^{(0)})}\,dW\label{appendix Z0 SDE}
\end{equation}
and with $loss$ shorthand for `the ultimate occurrence of loss', we have
\begin{equation}
\mathbf{E}_{y}\left[  q(Y_{\tau})\,|\,loss\right]  =\mathbf{E}_{y}\left[
q(Z_{\tau}^{(0)})\right]  .\label{mod E 0}
\end{equation}

Equations (\ref{appendix Z1 SDE}) and (\ref{appendix Z0 SDE}) are the
equations that the modified frequencies obey, while Eqs. (\ref{mod E 1}) and
(\ref{mod E 0}) contain the relations between conditioned and unconditioned expectations.

There are two features of the above results that are worth noting.

\noindent (1) The modified frequencies, $Z_{\tau}^{(0)}$ and $Z_{\tau}^{(1)}$,
are both independent of the sign of the selection coefficient, $\sigma$. In the 
modified picture, only the probabilities of fixation and loss have $\sigma$ dependence. 

\noindent (2) Full knowledge of $Z_{\tau}^{(1)}$ is sufficient to determine all properties of $Z_{\tau}^{(0)}$, 
since it can be shown that $\mathbf{E}_{y}\bigl[q(Z_{\tau}^{(0)}
)\bigr]=\mathbf{E}_{1-y}\bigl[q(1-Z_{\tau}^{(1)})\bigr]$ for arbitrary smooth $q(x)$.

\section{Small-noise approximation}

%%%%%%%%%%%%%%%%%%%%%%%%%%%%%%%%%%%%%%%%%%%%%%%%%%%%%%%%%%%%%%%%%%%%%%%%%%%%%%

\label{small-noise expansion appendix}

In this appendix we give details of the small-noise expansion \citep{Gardiner2009}
and the approximation that results from this.

We shall often refer to the random quantity $dW_{\tau}$ as \textit{noise}.

We proceed by: (i) carrying out a small-noise expansion of the random frequency,
$Z_{\tau}^{(a)}$; (ii) calculating important statistics associated with $Z_{\tau
}^{(a)}$; (iii) applying a small-noise approximation to the basic statistic $Q(\tau)$.

\subsection[Expansion of Ztaua]{Expansion of $Z_{\tau}^{(a)}$}
\label{Appendix subsection expansion}

We begin with the equation that the modified frequency $Z_{\tau}^{(a)}$ obeys,
namely Eq. (\ref{Z^a SDE}) of the main text:
\begin{equation}
dZ_{\tau}^{(a)}=f^{(a)}\left(  Z_{\tau}^{(a)}\right)  \,d\tau+\sqrt
{\varepsilon}\cdot g\left(  Z_{\tau}^{(a)}\right)  \,dW_{\tau}
\label{Z^a SDE appendix}
\end{equation}
where $f^{(a)}\left(  Z_{\tau}^{(a)}\right)$ is
given in Eq. (\ref{ff}) of the main text and
\begin{equation}
    g(z)=\sqrt{z(1-z)}.                         \label{g appendix}       
\end{equation}

Using $\varepsilon=1/R$ as a formal device to mark the strength of the noise
term $\sqrt{\varepsilon}\cdot g\left(  Z_{\tau}^{(a)}\right)  \,dW_{\tau}$ in
Eq. (\ref{Z^a SDE appendix}), we expand $Z_{\tau}^{(a)}$ as a power series in
$\sqrt{\varepsilon}$, writing
\begin{equation}
Z_{\tau}^{(a)}=Z_{\tau,0}^{(a)}+\sqrt{\varepsilon}Z_{\tau,1}^{(a)}+\varepsilon
Z_{\tau,2}^{(a)}+O(\varepsilon^{3/2}).\label{Za expand appendix}
\end{equation}
The coefficients of different powers of $\sqrt{\varepsilon}$ are treated as
being independent of $\varepsilon$. This allows us, on substituting the series
of Eq. (\ref{Za expand appendix}) into Eq. (\ref{Z^a SDE appendix} ), to
equate coefficients of like powers of $\varepsilon$ on the left and right-hand
sides.

On using $\varepsilon$ as a formal parameter that characterises
the noise, we treat it as independent of $R$. The selection term
$f^{(a)}\left(Z_{\tau}^{(a)}\right)$ in Eq. (\ref{Z^a SDE appendix}) is
not multiplied by noise, and thus its $R$ dependence is held exactly as
written and is not expressed in terms of $\varepsilon$. This treatment 
means the small-noise expansion differs from an asymptotic expansion in $R$,
where all $R$ dependence is a target of the expansion.

To make the
calculations manageable, we truncate the series in Eq. (\ref{Za expand appendix}) 
at order $\varepsilon$ to obtain the approximation
\begin{equation}
Z_{\tau}^{(a)}\simeq Z_{\tau,0}^{(a)}+\sqrt{\varepsilon}Z_{\tau,1}
^{(a)}+\varepsilon Z_{\tau,2}^{(a)}.\label{Za expand trunc appendix}
\end{equation}
We find
\begin{align}
dZ_{\tau,0}^{(a)} &  =f^{(a)}\left(  Z_{\tau,0}^{(a)}\right)  \,d\tau
\label{d Z0 ode appendix}\\
dZ_{\tau,1}^{(a)} &  =f^{(a)\,\prime}\left(  Z_{\tau,0}^{(a)}\right)
Z_{\tau,1}^{(a)}\,d\tau+g\left(  Z_{\tau,0}^{(a)}\right)  \,dW_{\tau
}\label{d Z1 ode appendix}\\
dZ_{\tau,2}^{(a)} &  =\left[  f^{(a)\,\prime}\left(  Z_{\tau,0}^{(a)}\right)
Z_{\tau,2}^{(a)}+\frac{1}{2}f^{(a)\,\prime\,\prime}\left(  Z_{\tau,0}
^{(a)}\right)  \left(  Z_{\tau,1}^{(a)}\right)  ^{2}\right]  \,d\tau
+g^{\,\prime}\left(  Z_{\tau,0}^{(a)}\right)  Z_{\tau,1}^{(a)}\,dW_{\tau
}.   \label{d Z2 ode appendix}
\end{align}
where a prime, $^{\prime}$, denotes differentiation of a function with respect
to its argument.

We see from Eq. (\ref{d Z0 ode appendix}) that $Z_{\tau,0}^{(a)}$ obeys an
equation without noise (it contains no $dW_{\tau}$ term) and we
shall often describe it as being \textit{deterministic}. Furthermore, the
initial value, $y$ is also independent of noise, and we  take $Z_{\tau
,0}^{(a)}$ in Eq.~(\ref{d Z0 ode appendix}) to be solely responsible for the
initial value, setting
\begin{equation}
Z_{0,0}^{(a)}=y.
\end{equation}
By contrast to $Z_{\tau,0}^{(a)}$, the terms $Z_{\tau,1}^{(a)}$ and
$Z_{\tau,2}^{(a)}$ contain noise in the equations they appear in (Eqs.
(\ref{d Z1 ode appendix}) and (\ref{d Z2 ode appendix})). We take all such
terms to be subject to an initial value of $0$:
\begin{equation}
Z_{0,1}^{(a)}=0\qquad Z_{0,2}^{(a)}=0.
\end{equation}

We shall use the approximation in Eq. (\ref{Za expand trunc appendix}) within
expectations. In this work we  determine the expectations to order
$\varepsilon$. After calculating an expectation we substitute
$\varepsilon=1/R$ into the final result, to obtain a small-noise approximation
of the expectation.

\subsection[Statistics associated with Ztaua]{Statistics associated with $Z_{\tau}^{(a)}$}

We shall shortly need some statistics associated with the random frequency
$Z_{\tau}^{(a)}$. In anticipation of this we give a method of their determination.

First, we note that the coefficient of $\sqrt{\varepsilon}$ in Eq.
(\ref{Za expand trunc appendix}), namely $Z_{\tau,1}^{(a)}$, has an expected
value of $0$, as directly follows from Eq. (\ref{d Z1 ode appendix}). Thus
\begin{equation}
\mathbf{E}_{y}\big[Z_{\tau,1}^{(a)}\big]=0.\label{Z1=0}
\end{equation}
Apart from $Z_{\tau,0}^{(a)}$, which is the noise free (or deterministic) part
of the random allele frequency $Z_{\tau}^{(a)}$, we shall also require the
statistics
\begin{equation}
M_{\tau}^{(a)}=\mathbf{E}_{y}\big[Z_{\tau,2}^{(a)}\big]\text{ and }S_{\tau
}^{(a)}=\mathbf{E}_{y}\Big[\left(  Z_{\tau,1}^{(a)}\right)  ^{2}
\Big].\label{M    S}
\end{equation}
We shall determine a set of coupled differential equations whose solution is
$Z_{\tau,0}^{(a)}$, $M_{\tau}^{(a)}$ and $S_{\tau}^{(a)}$. 

To proceed, we leave Eq. (\ref{d Z0 ode appendix}) unaltered. For $M_{\tau}^{(a)}$ we
directly take an expected value of Eq. (\ref{d Z2 ode appendix}), while for
$S_{\tau}^{(a)}$ we determine $d\left(  Z_{\tau,1}^{(a)}\right)  ^{2}$ from
Eq. (\ref{d Z1 ode appendix}) using It\^{o} calculus \citep{Tuckwell}, and then
take an expected value. We obtain
\begin{equation}
\left.
\begin{array}
[c]{rcl}
\dfrac{dZ_{\tau,0}^{(a)}}{d\tau} & = & f^{(a)}(Z_{\tau,0}^{(a)})\\
\dfrac{dM_{\tau}^{(a)}}{d\tau} & = & f^{(a)\,\prime}(Z_{\tau,0}^{(a)})\cdot
M_{\tau}^{(a)}+\frac{1}{2}f^{(a)\,\prime\,\prime}(Z_{\tau,0}^{(a)})\cdot
S_{\tau}^{(a)}\\
\dfrac{dS_{\tau}^{(a)}}{d\tau} & = & 2f^{(a)\,\prime}(Z_{\tau,0}^{(a)})\cdot
S_{\tau}^{(a)}+\big[g(Z_{\tau,0}^{(a)})\big]^{2}.
\end{array}
\right\}  \label{system}
\end{equation}
At $\tau=0$ the solutions take the values
\begin{equation}
Z_{0,0}^{(a)}=y,\qquad S_{0}^{(a)}=0,\qquad M_{0}^{(a)}=0.\label{ic}
\end{equation}
Equations (\ref{system}), subject to (\ref{ic}), are in a form amenable to
numerical solution from $\tau=0$ onwards. In Appendix \ref{integral representation appendix} an alternative
representation of the solution in terms of integrals is given.

\subsection[small-noise approximation of Q]{small-noise approximation of $Q(\tau)$}

To obtain a small-noise approximation of the basic statistic $Q(\tau)$ we make
use of the representation in Eq.~(\ref{Representation}):
\begin{equation}
Q(\tau)=\mathbf{E}_{y}\big[q(Y_{\tau})\big]=\mathbf{E}_{y}\big[q(Z_{\tau
}^{(0)})\big]\cdot\pi^{(0)}(y)+\mathbf{E}_{y}\big[q(Z_{\tau}^{(1)}
)\big]\cdot\pi^{(1)}(y)\label{Representation appendix}
\end{equation}
and the $Z_{\tau}^{(a)}$ from Eq.~(\ref{Za expand trunc appendix}). Keeping terms to order
$\varepsilon$ we have
\begin{align}
q(Z_{\tau}^{(a)}) &  \simeq q\left(  Z_{\tau,0}^{(a)}+\sqrt{\varepsilon
}Z_{\tau,1}^{(a)}+\varepsilon Z_{\tau,2}^{(a)}\right)  \nonumber\\
& \nonumber\\
&  \simeq q\left(  Z_{\tau,0}^{(a)}\right)  +\left(  \sqrt{\varepsilon}
Z_{\tau,1}^{(a)}+\varepsilon Z_{\tau,2}^{(a)}\right)  q^{\prime}\left(
Z_{\tau,0}^{(a)}\right)  +\frac{1}{2}\varepsilon\left(  Z_{\tau,1}
^{(a)}\right)  ^{2}q^{\prime\prime}\left(  Z_{\tau,0}^{(a)}\right)  .
\end{align}
Taking the expected value, and using Eqs. (\ref{Z1=0}) and (\ref{M    S}) we
obtain
\begin{align}
\mathbf{E}_{y}\big[q(Z_{\tau}^{(a)})\big]  & \simeq q\left(  Z_{\tau,0}
^{(a)}\right)  +\varepsilon q^{\prime}\left(  Z_{\tau,0}^{(a)}\right)
\mathbf{E}_{y}\big[Z_{\tau,2}^{(a)}\big]+\frac{1}{2}\varepsilon q^{\prime\prime
}(Z_{\tau,0}^{(a)})\mathbf{E}_{y}\big[\left(  Z_{\tau,1}^{(a)}\right)
^{2}\big]\nonumber\\
& \nonumber\\
& =q\left(  Z_{\tau,0}^{(a)}\right)  +\varepsilon q^{\prime}(Z_{\tau,0}
^{(a)})M_{\tau}^{(a)}+\frac{1}{2}\varepsilon q^{\prime\prime}(Z_{\tau,0}
^{(a)})S_{\tau}^{(a)}\nonumber\\
& \nonumber\\
& =q\left(  Z_{\tau,0}^{(a)}\right)  +\frac{1}{R}\left[  q^{\prime}(Z_{\tau
,0}^{(a)})M_{\tau}^{(a)}+\frac{1}{2}q^{\prime\prime}(Z_{\tau,0}^{(a)})S_{\tau}
^{(a)}\right]  .
\end{align}
Using this result in Eq. (\ref{Representation appendix}) yields
\begin{equation}
Q(\tau)\simeq Q_{0}(\tau)+\frac{1}{R}Q_{1}(\tau)\label{Q approx appendix}
\end{equation}
where
\begin{equation}
Q_{0}(\tau)=\sum_{a=0}^{1}q(Z_{\tau,0}^{(a)})\cdot\pi^{(a)}(y)\equiv
\left\langle q(Z_{\tau,0})\right\rangle _{\pi}\label{Q0}
\end{equation}
and
\begin{align}
Q_{1}(\tau)  & =\sum_{a=0}^{1}q^{\,\prime}(Z_{\tau,0}^{(a)})\cdot M_{\tau
}^{(a)}\cdot\pi^{(a)}(y)+\sum_{a=0}^{1}\frac{1}{2}q^{\,\prime\,\prime}
(Z_{\tau,0}^{(a)})\cdot S_{\tau}^{(a)}\cdot\pi^{(a)}(y)\nonumber\\
& \nonumber\\
& \equiv\left\langle q^{\,\prime}(Z_{\tau,0})\cdot M_{\tau}\right\rangle
_{\pi}+\frac{1}{2}\left\langle q^{\,\prime\,\prime}(Z_{\tau,0})\cdot S_{\tau
}\right\rangle _{\pi}\label{Q1}
\end{align}
With  $Z_{\tau,0}^{(a)}$, $M_{\tau}^{(a)}$ and $S_{\tau}^{(a)}$ determined
from Eqs. (\ref{system}) and (\ref{ic}),  Eq. (\ref{Q approx appendix})
constitutes a small-noise approximation of $Q(\tau)=\mathbf{E}_{y}
\big[q(Y_{\tau})\big]$.

\section{Integral representations of trajectory statistics}

\label{integral representation appendix}

In this appendix we determine integral representations of $S_{\tau}
^{(a)}=E_{y}\left[  \left(  Z_{\tau,1}^{(a)}\right)  ^{2}\right]  $ and
$M_{\tau}^{(a)}=E_{y}\left[  Z_{\tau,2}^{(a)}\right]  $ in terms of the
deterministic trajectory, $Z_{\tau,0}^{(a)}$. We express the results in terms
of the functions $f^{(a)}(z)$ of Eq. (\ref{ff}) of the main text and $g(z)$ of
Eq. (\ref{g appendix}),
and use a prime, $^{\prime}$, to denote differentiation of a function with
respect to its argument.

\subsection{Derivation}

The equations found for the coefficients $Z_{\tau,0}^{(a)}$, $Z_{\tau,1}
^{(a)}$ and $Z_{\tau,2}^{(a)}$ in the expansion of $Z_{\tau}^{(a)}$ are given
by Eqs. (\ref{d Z0 ode appendix}), (\ref{d Z1 ode appendix}) and (\ref{d Z2 ode appendix}):
\begin{align}
dZ_{\tau,0}^{(a)}  &  =f^{(a)}\left(  Z_{\tau,0}^{(a)}\right)  \,d\tau
\label{d Z0 ode appendix 2}\\
dZ_{\tau,1}^{(a)}  &  =f^{(a)\,\prime}\left(  Z_{\tau,0}^{(a)}\right)
Z_{\tau,1}^{(a)}\,d\tau+g\left(  Z_{\tau,0}^{(a)}\right)  \,dW_{\tau
}\label{d Z1 ode appendix 2}\\
dZ_{\tau,2}^{(a)}  &  =\left[  f^{(a)\,\prime}\left(  Z_{\tau,0}^{(a)}\right)
Z_{\tau,2}^{(a)}+\frac{1}{2}f^{(a)\,\prime\,\prime}\left(  Z_{\tau,0}
^{(a)}\right)  \left(  Z_{\tau,1}^{(a)}\right)  ^{2}\right]  \,d\tau
+g^{\,\prime}\left(  Z_{\tau,0}^{(a)}\right)  Z_{\tau,1}^{(a)}\,dW_{\tau}.
\label{d Z2 ode appendix 2}
\end{align}

We first derive the implicit equation that determines $Z_{\tau,0}^{(a)}$ and
shall then obtain integral representations of $Z_{\tau,1}^{(a)}$ and
$Z_{\tau,2}^{(a)}$ in terms of $Z_{\tau,0}^{(a)}$.

\subsection[Equation for Ztaua]{Equation for $Z_{\tau,0}^{(a)}$}

The equation obeyed by $Z_{\tau,0}^{(a)}$ is given by Eq.
(\ref{d Z0 ode appendix 2}) and is subject to
\begin{equation}
Z_{0,0}^{(a)}=y. \label{Z0=y appendix}
\end{equation}
From Eqs. (\ref{d Z0 ode appendix}) and (\ref{Z0=y appendix}) we obtain
\begin{equation}
\tau=\int_{y}^{Z_{\tau,0}^{(a)}}\frac{dz}{f^{(a)}\left(  z\right)  }.
\label{implicit eq for Z}
\end{equation}
This equation implicitly determines $Z_{\tau,0}^{(a)}$ and given the known
form for $f^{(a)}\left(  z\right)  $ can be used to determine properties of
$Z_{\tau,0}^{(a)}$. Equation (\ref{implicit eq for Z}) could be a starting
point for an analytic approximation of $Z_{\tau,0}^{(a)}$ but we will not
pursue this here.

\subsection[Equation for Ztau1a]{Equation for $Z_{\tau,1}^{(a)}$ and its first and second moment}

Assuming knowledge of $Z_{\tau,0}^{(a)}$ we now wish to determine $Z_{\tau
,1}^{(a)}$ which is subject to $Z_{0,1}^{(a)}=0$.

We can write Eq. (\ref{d Z1 ode appendix 2}) as
\begin{equation}
d\left[  \exp\left(  -\int_{0}^{\tau}f^{(a)\,\prime}\left(  Z_{\sigma,0}
^{(a)}\right)  d\sigma\right)  Z_{\tau,1}^{(a)}\right]  \,=\exp\left(
-\int_{0}^{\tau}f^{(a)\,\prime}\left(  Z_{\sigma,0}^{(a)}\right)
d\sigma\right)  g\left(  Z_{\tau,0}^{(a)}\right)  \,dW_{\tau}
\end{equation}
hence
\begin{equation}
Z_{\tau,1}^{(a)}=\exp\left(  \int_{0}^{\tau}f^{(a)\,\prime}\left(
Z_{\sigma,0}^{(a)}\right)  d\sigma\right)  \int_{0}^{\tau}\exp\left(
-\int_{0}^{\mu}f^{(a)\,\prime}\left(  Z_{\sigma,0}^{(a)}\right)
d\sigma\right)  g\left(  Z_{\mu,0}^{(a)}\right)  \,dW_{\mu}. \label{Z1}
\end{equation}
This can be written in a simpler form. Noting that $dZ_{\tau,0}^{(a)}
=f^{(a)}\left(  Z_{\tau,0}^{(a)}\right)  \,d\tau$ we can write
\begin{align}
\int_{0}^{\tau}f^{(a)\,\prime}\left(  Z_{\sigma,0}^{(a)}\right)  d\sigma &
=\int_{0}^{\tau}\frac{f^{(a)\,\prime}\left(  Z_{\sigma,0}^{(a)}\right)
}{f^{(a)}\left(  Z_{\sigma,0}^{(a)}\right)  }f^{(a)}\left(  Z_{\sigma,0}
^{(a)}\right)  d\sigma\nonumber\\
&  =\int_{y}^{Z_{\tau,0}^{(a)}}\frac{f^{(a)\,\prime}\left(  z\right)
}{f^{(a)}\left(  z\right)  }dz=\ln\left(  \frac{f^{(a)}\left(  Z_{\tau
,0}^{(a)}\right)  }{f^{(a)}(y)}\right)  .
\end{align}
Using this result in Eq. (\ref{Z1}) yields
\begin{equation}
Z_{\tau,1}^{(a)}=f^{(a)}\left(  Z_{\tau,0}^{(a)}\right)  \int_{0}^{\tau}
\frac{g\left(  Z_{\sigma,0}^{(a)}\right)  }{f^{(a)}\left(  Z_{\sigma,0}
^{(a)}\right)  }\,dW_{\sigma}. \label{Z1a rep}
\end{equation}
This constitutes the solution of $Z_{\tau,1}^{(a)}$. Note that because
$E[dW_{\sigma}]=0$ we have
\begin{equation}
E_{y}\left[  Z_{\tau,1}^{(a)}\right]  =0.
\end{equation}
Squaring and averaging Eq. (\ref{Z1a rep}) yields
\begin{align}
E_{y}\left[  \left(  Z_{\tau,1}^{(a)}\right)  ^{2}\right]   &  =\left[
f^{(a)}\left(  Z_{\tau,0}^{(a)}\right)  \right]  ^{2}\int_{0}^{\tau}
\frac{\left[  g\left(  Z_{\sigma,0}^{(a)}\right)  \right]  ^{2}}{\left[
f^{(a)}\left(  Z_{\sigma,0}^{(a)}\right)  \right]  ^{2}}\,d\sigma\nonumber\\
&  =\left[  f^{(a)}\left(  Z_{\tau,0}^{(a)}\right)  \right]  ^{2}\int
_{0}^{\tau}\frac{\left[  g\left(  Z_{\sigma,0}^{(a)}\right)  \right]  ^{2}
}{\left[  f^{(a)}\left(  Z_{\sigma,0}^{(a)}\right)  \right]  ^{3}}
\,f^{(a)}\left(  Z_{\sigma,0}^{(a)}\right)  d\sigma\nonumber\\
&  =\left[  f^{(a)}\left(  Z_{\tau,0}^{(a)}\right)  \right]  ^{2}\int
_{y}^{Z_{\tau,0}^{(a)}}\frac{\left[  g\left(  z\right)  \right]  ^{2}}{\left[
f^{(a)}\left(  z\right)  \right]  ^{3}}\,dz. \label{EZ1^2}
\end{align}
In the main text we define $S_{\tau}^{(a)}=E_{y}\left[  \left(  Z_{\tau
,1}^{(a)}\right)  ^{2}\right]  $ thus an integral representation of $S_{\tau
}^{(a)}$ is given by
\begin{equation}
S_{\tau}^{(a)}=\left[  f^{(a)}\left(  Z_{\tau,0}^{(a)}\right)  \right]
^{2}\int_{y}^{Z_{\tau,0}^{(a)}}\frac{\left[  g\left(  z\right)  \right]  ^{2}
}{\left[  f^{(a)}\left(  z\right)  \right]  ^{3}}\,dz.  \label{s integral}
\end{equation}

\subsection[Equation for Ztau2a]{Equation for $Z_{\tau,2}^{(a)}$ and its first moment}

We can write
\begin{equation}
dZ_{\tau,2}^{(a)}-f^{(a)\,\prime}\left(  Z_{\tau,0}^{(a)}\right)  Z_{\tau
,2}^{(a)}\,d\tau=\frac{1}{2}f^{(a)\,\prime\,\prime}\left(  Z_{\tau,0}^{(a)}\right)
\left(  Z_{\tau,1}^{(a)}\right)  ^{2}\,d\tau+g^{\,\prime}\left(  Z_{\tau
,0}^{(a)}\right)  Z_{\tau,1}^{(a)}\,dW_{\tau}
\end{equation}
and similar to the procedure for $Z_{\tau,1}^{(a)}$ we obtain
\begin{align}
Z_{\tau,2}^{(a)}  &  =f^{(a)}\left(  Z_{\tau,0}^{(a)}\right)  \int_{0}^{\tau
}\frac{\frac{1}{2}f^{(a)\,\prime\,\prime}\left(  Z_{\sigma,0}^{(a)}\right)
\left(  Z_{\sigma,1}^{(a)}\right)  ^{2}}{f^{(a)}\left(  Z_{\sigma,0}
^{(a)}\right)  }\,d\sigma\nonumber\\
&  \quad+f^{(a)}\left(  Z_{\tau,0}^{(a)}\right)  \int_{0}^{\tau}
\frac{g^{\,\prime}\left(  Z_{\sigma,0}^{(a)}\right)  Z_{\sigma,1}^{(a)}
}{f^{(a)}\left(  Z_{\sigma,0}^{(a)}\right)  }\,dW_{\sigma}.
\end{align}
This has a non-zero expectation
\begin{align}
E_{y}\left[  Z_{\tau,2}^{(a)}\right]   &  =\frac{1}{2}f^{(a)}\left(
Z_{\tau,0}^{(a)}\right)  \int_{0}^{\tau}\frac{f^{(a)\,\prime\,\prime}\left(
Z_{\sigma,0}^{(a)}\right)  }{f^{(a)}\left(  Z_{\sigma,0}^{(a)}\right)  }
E_{y}\left[  \left(  Z_{\sigma,1}^{(a)}\right)  ^{2}\right]  \,d\sigma
\nonumber\\
&  =\frac{1}{2}f^{(a)}\left(  Z_{\tau,0}^{(a)}\right)  \int_{0}^{\tau}
\frac{f^{(a)\,\prime\,\prime}\left(  Z_{\sigma,0}^{(a)}\right)  }
{f^{(a)}\left(  Z_{\sigma,0}^{(a)}\right)  }\left(  \left[  f^{(a)}\left(
Z_{\sigma,0}^{(a)}\right)  \right]  ^{2}\int_{y}^{Z_{\sigma,0}^{(a)}}
\frac{\left[  g\left(  z\right)  \right]  ^{2}}{\left[  f^{(a)}\left(
z\right)  \right]  ^{3}}\,dz\right)  d\sigma\nonumber\\
&  =\frac{1}{2}f^{(a)}\left(  Z_{\tau,0}^{(a)}\right)  \int_{y}^{Z_{\tau
,0}^{(a)}}du\int_{y}^{u}dv\,f^{(a)\,\prime\,\prime}\left(v\right)
\frac{\left[  g\left(  v\right)  \right]  ^{2}}{\left[  f^{(a)}\left(
v\right)  \right]  ^{3}}\,\nonumber\\
&  =\frac{1}{2}f^{(a)}\left(  Z_{\tau,0}^{(a)}\right)  \int_{y}^{Z_{\tau
,0}^{(a)}}\left[  \,f^{(a)\,\,\prime}\left(  Z_{\tau,0}^{(a)}\right)
-f^{(a)\,\,\prime}\left(v\right)  \right]  \frac{\left[  g\left(  v\right)
\right]  ^{2}}{\left[  f^{(a)}\left(  v\right)  \right]  ^{3}}\,dv.
\end{align}
In the main text we define $M_{\tau}^{(a)}=E_{y}\left[  Z_{\tau,2}
^{(a)}\right]  $ thus an integral representation of $M_{\tau}^{(a)}$ is given
by
\begin{equation}
M_{\tau}^{(a)}=\frac{1}{2}f^{(a)}\left(  Z_{\tau,0}^{(a)}\right)  \int
_{y}^{Z_{\tau,0}^{(a)}}\left[  \,f^{(a)\,\,\prime}\left(  Z_{\tau,0}
^{(a)}\right)  -f^{(a)\,\,\prime}\left(v\right)  \right]  \frac{\left[
g\left(  v\right)  \right]  ^{2}}{\left[  f^{(a)}\left(  v\right)  \right] 
^{3}}\,dv.    \label{M integral}
\end{equation}

\section{Estimating error}

\label{error appendix}

In this appendix, we motivate and specify how we quantify the error between
the exact result for the statistic $Q(\tau)$ and its approximation, which we
write here as $Q_{\text{approx}}(\tau)$.

The method used is tailored to the assumed property that $Q(\tau)\geq0$, and
that as $\tau$ becomes large, both $Q(\tau)$ and $Q_{\text{approx}}(\tau)$ tend
to the same asymptotic value, which we write as $Q(\infty)$. Note that we do
not assume $Q(\tau)$ behaves monotonically with $\tau$.

A natural measure of the error with which $Q_{\text{approx}}(\tau)$
approximates $Q(\tau)$ is the fractional error, $[Q_{\text{approx}}
(\tau)-Q(\tau)]/Q(\tau)$. While intuitive, its practical significance is
questionable. Suppose $Q(\tau)$ can take a range of values, say from $0$ to
$1$. A fractional error of $100\%$ could arise from $Q_{\text{approx}}
(\tau)=2\times10^{-6}$ while $Q(\tau)=10^{-6}$. Such an error would be flagged
as problematic, but in practice is likely to be considered of no consequence,
since the difference between $Q_{\text{approx}}(\tau)$ and $Q(\tau)$ is
extremely small.

Rather than asking \textquotedblleft how wrong the approximation is at each
$\tau$?\textquotedblright, we ask the more relevant question:
\textquotedblleft what fraction of the curve of $Q(\tau)$ (the exact result)
is misplaced or misrepresented by the approximation?\textquotedblright\ To
answer this question, we use $\tau$ in the range $0\leq\tau\leq\kappa$ that we
call the `active window' and define an error metric by
\begin{equation}
\Delta=\frac{\int_{0}^{\kappa}|Q_{\text{approx}}(\tau)-Q(\tau)|\,d\tau}
{\int_{0}^{\kappa}Q(\tau)\,d\tau}.
\end{equation}
To specify the cut-off time, $\kappa$, we proceed by computing the 
total integrated absolute deviation of the exact statistic:
\begin{equation}
D_{\text{total}}=\int_{0}^{\infty}|Q(\tau)-Q(\infty)|\,d\tau.
\label{Dtotal appendix}
\end{equation}
We then define $\kappa$ as the smallest time at which the integrated absolute deviation
reaches $99\%$ of its total value. That is:
\begin{equation}
\kappa=\min\left\{  \tau\;\middle|\;\int_{0}^{\tau}|Q(\sigma)-Q(\infty
)|\,d\sigma\geq0.99\times D_{\text{total}}\right\}  .
\end{equation}

In practice, we approximate $D_{\text{total}}$ in Eq. (\ref{Dtotal appendix})
by using a large but finite time, $\tau_{\max}$, as the upper limit, and we
also approximate $Q(\infty)$ by $Q(\tau_{\max})$.

With the above, no knowledge of $Q(\tau)$ outside the range $0\leq\tau\leq
\tau_{\max}$ is required to evaluate the error, $\Delta$, which applies to
both monotonic and non-monotonic curves, as well as the case where
$Q(\infty)=0$. The interval $0\leq\tau\leq\kappa$ represents the region where
$Q(\tau)$ is dynamically active, with the remaining tail contributing only
$1\%$ of the total absolute variation. By truncating the range of the
integrals in $\Delta$ at $\kappa$ we ensure that the denominator in $\Delta$
reflects only the main region where $Q(\tau)$ varies, and is not unjustifiably
inflated by using a larger upper limit to the integral, which can lead to a
smaller value of $\Delta$.

Note that $\Delta$ could be close to $100\%$ if $Q_{\text{approx}}(\tau)$ is
very small over the range $0\leq\tau\leq\kappa$. Additionally, $\Delta$ is not
inherently bounded by $100\%$. For example if over a wide range of $\tau$ we
have $Q_{\text{approx}}(\tau)>2Q(\tau)$, then $|Q_{\text{approx}}(\tau
)-Q(\tau)|>Q(\tau)$ over that range, and this may drive $\Delta$ above unity.
Generally, small values of $\Delta$ are indicative of a reasonable
approximation while large values suggest problems with the approximation.

\newpage

\bibliography{Refs}

\end{document}